# The Edges of Our Universe

Toby Ord[*]

This paper explores the fundamental causal limits on how much of the universe we can observe or affect. It distinguishes four principal regions: the *affectable* universe, the *observable* universe, the *eventually observable* universe, and the *ultimately observable* universe. It then shows how these (and other) causal limits set physical bounds on what spacefaring civilisations could achieve over the longterm future.

**Introduction**

How large is the universe? What exactly is the observable universe? Will we ever be able to detect things that are outside it? If so, what are the ultimate limits of observability? Are there fundamental limits on how far we could travel through space? How far away does something have to be such that it is completely causally separate from us? How do these relate to each other? And how do they change with time?

These are key questions for understanding the large scale picture of the universe, its spatial limits, and its causal structure. They are also key questions when attempting to understand how much a spacefaring civilisation might ever be able to achieve — including fundamental physical limits on our own civilisation. Yet very few people understand these limits and how they relate to each other. Many discussions conflate all the different causal limits together and assume that they are all set by the observable universe. Even astrophysicists and cosmologists make mistakes about these limits, including in their academic articles and textbooks.[1] Some of the most fundamental limits are rarely mentioned and for one important limit, I know of no scientific literature which mentions it at all.

In this paper, I explain and clarify these limits. Using spacetime diagrams, I distinguish nine different concentric shells around the Earth, billions of light years in diameter, showing what each of them means, how they evolve over time, and how they relate to one another. I then explore their meaning and importance by looking at how these different edges of the universe provide limits on what it is physically possible for an ambitious spacefaring civilisation to achieve. I finish with a convenient summary of each limit and its current value.

---

[*] Future of Humanity Institute, University of Oxford.

This paper owes a great debt to conversations with many people interested in these questions. In particular, I'd like to thank Stuart Armstrong, Nick Bostrom, Tamara Davis, Peter Eckersley, Jay Olson, Max Tegmark, Martin Rees, and especially, Anders Sandberg.

[1] Davis and Lineweaver (2003) list some prominent examples.

Surprisingly, the necessary cosmological equations do not involve mathematics beyond what is taught at high school.[2] This creates a rare opportunity for people from other disciplines, and even the general public, to learn about cosmologists' current understanding of the fundamental structure of the universe.[3] I have endeavoured to make the most of this opportunity, by presenting the real physics in a way that is widely accessible without compromising at all on its correctness.

Some of the more unexpected things we will see include:

- Many galaxies that are currently outside the observable universe will become observable later.
- Less than 5% of the galaxies we can currently observe could ever be affected by us, and this is shrinking all the time.
- But we *can* affect some of the galaxies that are receding from us faster than the speed of light.
- There is a fundamental split in the longterm history of the universe in 150 billion years' time between an era of connection and an era of isolation.

I shall examine the edges of our universe as they occur under the most widely accepted cosmological model (ΛCDM), in which the expansion of space continues to accelerate due to a cosmological constant. There has been substantial experimental support for this model over the last two decades, with its parameters being estimated to 1% accuracy. But this is not settled science and there is some evidence that casts doubt on it. If it is replaced with something substantially different — in particular, if the replacement does not feature accelerating expansion of space — then the limits I describe may not in fact govern our universe and a new analysis would need to be done. (I explain more about this in the final section.)

While I shall describe finite spherical parts of the universe that are of particular interest to us, I do not mean to imply that any of these is the entire universe — the entirety of spacetime. Indeed, since they are all centred on us, this would be suspiciously anthropocentric. Cosmologists generally believe that outside all of these spheres, the pattern of galaxies continues much the same as within. They also strongly suspect that the entire universe does not have an edge. It may be finite (wrapping around on itself like the surface of a sphere) or it may be infinite. Even if it is finite, evidence strongly suggests it is much larger than the regions we shall discuss. However, according to physical laws as we currently understand them, the parts of the universe that are causally connected to us in various ways are all finite and all have a (spherical) edge.

---

[2] In their full generality, they would require postgraduate level mathematics. But as applied to the fundamental distance limits in the universe as we currently understand it, one can get by with nothing more than square roots, integrals, and use of a spreadsheet.

[3] Indeed, an interested layperson could use the equation and spreadsheet method from the Appendix to calculate the age of the universe, the size of the observable universe, and many other things directly from the five measured constants given.

In Part One, I classify and clarify these parts of the universe.

In Part Two, I explore the limits they impose on interstellar civilisation.

**PART ONE — THE PHYSICS**

**Spacetime diagrams**

A spacetime diagram is a tool for understanding motion and the limits of causality. It is a graph with location in space on the horizontal axis and time on the vertical axis. While location in space is three dimensional (so would ideally have three axes of its own), it is common practice to just show a single spatial dimension. This usually suffices to show the phenomena of interest, and is much easier represent on paper.

Spacetime diagrams were originally created in the context of special relativity and its simple flat spacetime. We shall review the basics of these diagrams in that context, before showing how they can be modified to represent our expanding universe.

The trajectory of an object over time is represented by a path on the diagram known as a *world line*. An object that is stationary with respect to the coordinate frame is represented by a path that has the same spatial co-ordinate at all times — a vertical line. The grey lines in *Figure 1* thus represent evenly spaced objects which are at rest. An object that moves at constant speed has a straight diagonal path, while an accelerating object has a curved path. Since light moves at a constant speed ($c$), light rays have straight diagonal paths. As this speed is so important, the scale of the spacetime diagram is usually set so that light's path is at a 45° angle. For this reason, and others, it is convenient to use units where the speed of light is equal to 1. We will measure time in years and the space in light-years.

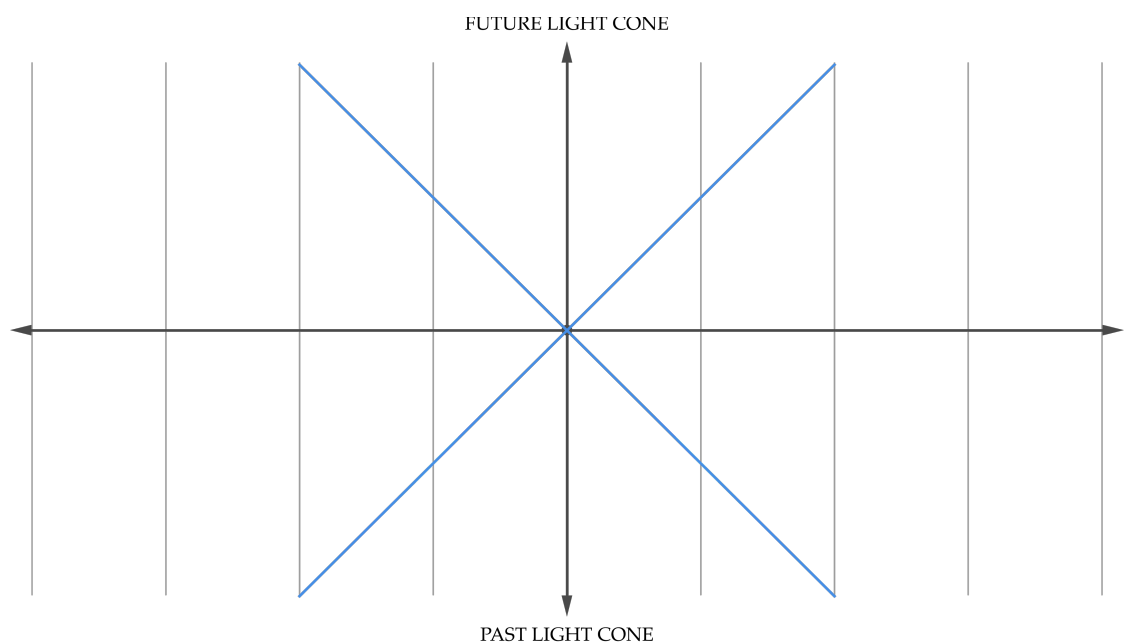

*Figure 1.* A spacetime diagram for special relativity.

If we are interested in a particular point in space and time (an *event*), it can be helpful to draw light rays emanating from that point and light rays converging to that point. The region between the light rays that emanate from an event is called its *future light cone* and represents the region of spacetime that could be reached from the event travelling at the speed of light or slower. This corresponds to every location in spacetime that is causally affectable by that event. The region between the light rays that converge to the event is the *past light cone* and comprises all points in spacetime from which the event could be reached at the speed of light or slower. This corresponds to every location in spacetime that could causally affect the event (and thus to every location that could be observed by the event).

If we consider any pair of events (the two dots in *Figure 2*) and examine their light cones, we will see that there is always a region in the future where their future light cones overlap. This corresponds to the points in space and time where the events could interact with each other. Its lowest point is the first time and place at which they could interact (if signals were sent towards each other at the speed of light). The events will also always have overlapping past light cones, which correspond to prior events that could have influenced both the marked events.

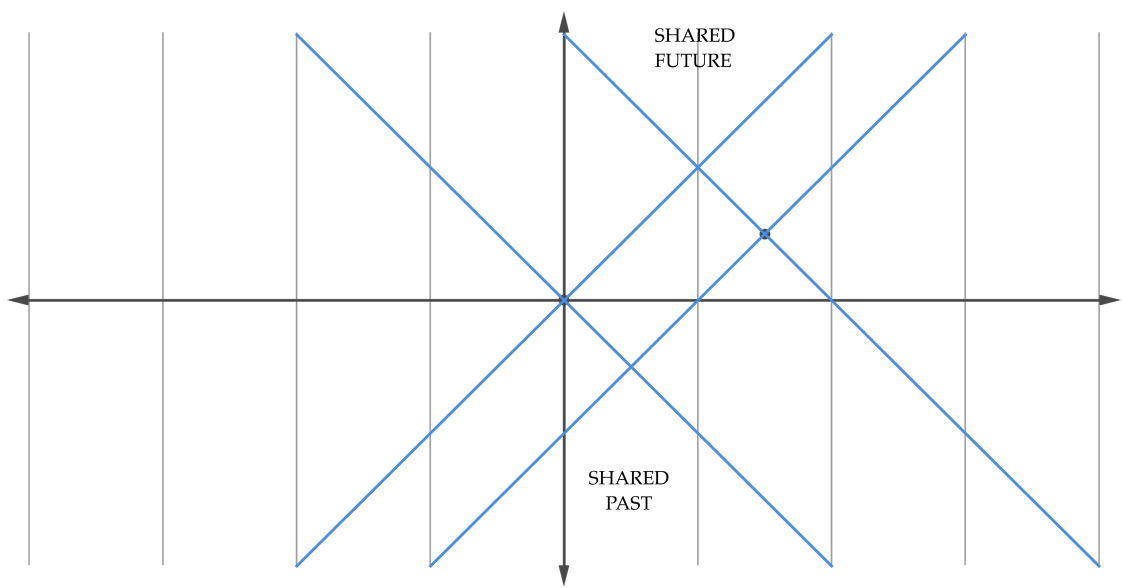

*Figure 2.* Overlapping light cones.

**Our universe**

Our universe differs from the picture above in fundamental ways. It is curved rather than flat, it expands over time, and it has a finite history. The first of these is very relevant in locations near exceptionally dense objects, but turns out not to make much difference for our purposes, as space is actually very flat on the large scales we shall consider. The other two change the structure of spacetime and (and spacetime diagrams) in ways that are crucial for understanding the universe on the largest scales.

If we look at distant galaxies, we can see that they are receding from us and from each other. They do this is in a very regular manner — galaxies *k* times further away are receding *k* times faster. Our best understanding of this is that the space between them is expanding, so the galaxies are like marks drawn on a rubber band which is stretching over time.[4] If we took our measurements and extrapolated backwards in time, we would see space becoming smaller and smaller and the galaxies moving closer together. If we go far enough, we find that around 13.8 billion years ago, space and all the matter in it converge to become extremely small, extremely dense, and extremely hot — an event known as the Big Bang. We don't fully understand this event and understand even less about its origins. For our purposes, we take the Big Bang (and the zero point of our time scale) to be the point at which the universe began expanding in its current fashion. We thus set aside a hypothetical period of much more rapid expansion that may have preceded this, as posited by the theory of *inflation*.

The rate of expansion of space depends upon what is in it, and thus obeys a fairly complicated equation. The scale of space at given time relative to its present scale is called the *scale factor* and is denoted $a(t)$. Since space was smaller in the past, values in the past are less than 1, with $a(t)$ approaching 0 at the Big Bang. We don't entirely know what will happen to the scale factor in the future, but the most widely accepted cosmological model ($\Lambda$CDM) suggests that the future evolution of the universe is dominated by 'dark energy'. This implies that $a(t)$ will continue to grow without bound, and as time goes on it will approach an exponential rate of growth, with lengths doubling every 12 billion years. For the purposes of this paper, we assume that this prediction is correct — that the universe's rate of expansion will increase exponentially — and see what this means for understanding our cosmic limits. The mathematical details for calculating $a(t)$ are given in the appendix.

The scale factor does not make *all* distances bigger. If a collection of particles are bound together by a force (such as gravity), they will remain the same distance apart that they currently are, even as space stretches. A reasonable analogy would be two masses connected with a spring, sitting on a rubber sheet. A strong enough spring will keep the objects at a constant distance, even if the sheet is being stretched exponentially.[5] Thus, an atom, a planet, a galaxy, and a galactic cluster (each of which is bound) will not grow in line with the scale factor.

However, the scale factor does increase the distances *between* the gravitationally bound collections of galaxies. Our galaxy is bound only to the other 54 members of our Local Group,[6] so at a future time *t* all galaxies outside our group will be $a(t)$ times further away from us than they currently are. The largest bound structures are galactic clusters which have radii of about 10 million light years. But since the distances we shall be exploring in this paper are on the order of *billions* of light years, they will increase with the scale factor.

---

[4] An excellent explanation of the expansion of space — clarifying many misconceptions — can be found in Lineweaver & Davis (2005).

[5] Note that the objects will be slightly further apart than they would be in a non-expanding universe.

[6] (Nagamine & Loeb 2003), (Busha et al 2003).

As galaxies get further apart, it takes light longer to travel between them. Eventually space will expand so quickly that light cannot travel the ever-expanding gulf between our Local Group and its nearest neighbouring group (simulations suggest that this will take around 150 billion years).[7] Since nothing is faster than light, there will be no way to reach neighbouring groups and our Local Group will be causally cut off from affecting the rest of the universe. Indeed all such groups will become isolated at a similar time.

So far, the way we have been talking of distance is *proper distance*. The proper distance between groups of galaxies is expanding in line with the scale factor. An alternative measure of distance is *comoving distance*. This is the proper distance divided by the scale factor. The comoving distance between groups of galaxies is therefore roughly constant (changed only by their modest amount of relative motion, not by the expansion of space). You can think of comoving coordinates roughly as using the distribution of galaxies in space as a grid on which distances are measured. Comoving distance is very useful for reasoning about the very large scales of time and space that we are interested in as one doesn't have to keep adjusting for the changing scale of space.

**Diagrams for our universe**

What does a spacetime diagram look like for our universe, with a finite past and expanding space? *Figure 3* shows the future light cone of an event at time zero (at the Big Bang[8]). The horizontal axis now shows space in comoving coordinates. The dot on the vertical axis is our current time and location. The vertical grey lines are the world lines of comoving galaxies.

The two light rays emanating from the event are curved on this diagram due to the expansion of space. The slope increases over time because light takes longer and longer to travel each light-year of comoving distance as time goes on — relative to the background grid of galaxies, the light is decelerating. Due to the accelerating expansion of our universe, which will eventually result in exponential expansion, these curves have vertical asymptotes.[9] These represent a bound in comoving distance beyond which light from the initial event can never travel. This is 62.9 billion light years. In other words, if a photon were released from here at the time of the Big Bang, the furthest galaxy it could ever reach (in the infinite limit of time) is a galaxy that is currently 62.9 billion light years away. Because it began its journey at the earliest possible time, this is the furthest a ray of light can travel and is a fundamental property of our universe. Things that are more than 62.9 billion light years apart (in comoving distance) cannot affect each other — no matter how early they start moving — because nothing can travel from one to the other.

---

[7] (Nagamine & Loeb 2003) give a rough estimate of 100 billion years. Busha et al (2003) calculate 175 billion years. My own calculations with the 2015 figures for dark energy give a value of around 150 billion years.

[8] Or, if the theory of inflation is true, then my time zero represents the time at which inflation ceased.

[9] To see what would happen if the expansion did not continue to accelerate, see the section *What if ΛCDM is wrong?* near the end of Part 2.

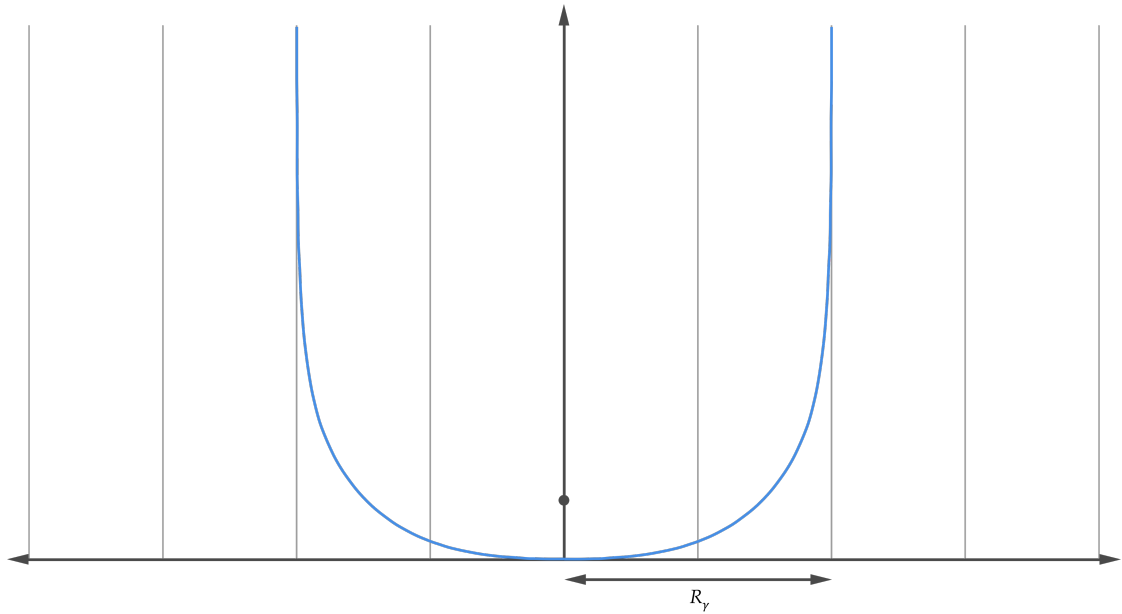

*Figure 3.* A spacetime diagram for our expanding universe. The horizontal axis shows space in comoving coordinates. The curves describe the future light cone for an event at the Big Bang.

This shape of the light rays is also a fundamental property of our expanding universe. The path of any light ray on a diagram like this — emitted at any time and place — will be a piece of this shape. If it is emitted from a different point in space, it will be translated to the left or right. If it is emitted in the opposite direction, it will be reflected (as the two rays on this diagram are reflections of each other). If it is emitted at a different time, it will *not* be translated upwards, but will instead be translated sideways (as we will see in the next diagram), and will begin part way along the curve. If it is absorbed, it will end at the time of absorption. The fact that all rays of light have this shape on this diagram is what makes it extremely useful for exploring the fundamental causal limits of our universe.

We shall denote the distance light can travel by time $t$ as $d_\gamma(t)$ and the limit of this as time approaches infinity as $D_\gamma$. As explained above, $D_\gamma$ is the furthest light can travel and is approximately 62.9 billion light years. The equation for $d_\gamma(t)$ is given in the appendix.

The region between the two rays of light is the future light cone of the event, representing all the points in spacetime that could be causally affected by this event. Because the event is at time zero, this is the largest possible future light cone — the largest region of spacetime that can be affected starting at one point in spacetime.

*The affectable universe*

What do our own light cones look like? *Figure 4* shows the past and future light cones for an observer at our time — 13.8 billion years after the Big Bang. The lines that make it up have exactly the same shape as before, just shifted horizontally towards each other, until the point where they cross moves up to 13.8 billion years. Our future light cone is much smaller than that in *Figure 3*. Its eventual radius has shrunk to about a quarter of its former size (from 62.9 billion light years to 16.5 billion light years) making its eventual volume shrink to about 2% of what it once included.

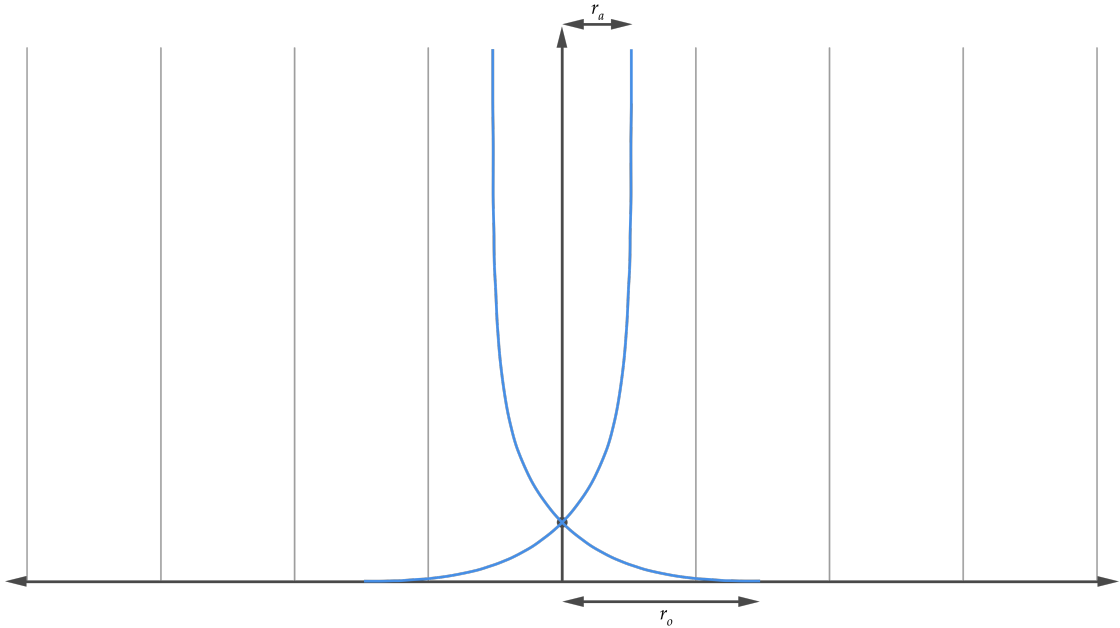

*Figure 4.* The past and future light cones for Earth, now. Their top and bottom are the current affectable universe and observable universe.

This sphere of radius 16.5 billion light years, centred on the Earth, is one of the answers to where the edge of our universe lies. Everything inside it is causally affectable by us. Indeed, the effects of some of our actions will reach the edge itself. Photons transmitted from the Earth today in our television broadcasts will take the path of the upper blue lines on *Figure 4*, converging on this distance (though they will be so thinly distributed and so redshifted that they might be undetectable in practice). However, everything outside this sphere is forever beyond our causal reach. Cosmologists refer to this spherical boundary as the universe's event horizon (from an analogy with the event horizon of a black hole).

Note that by symmetry, this distance we could eventually reach with a light signal starting here and now is equal to the distance from which an event in another galaxy happening now could eventually affect us here. Indeed this alternate perspective is how cosmologists typically define this boundary. In some sense these are equally good definitions as they will always give the same answer, but I feel that the cosmologists' standard definition doesn't highlight quite how important this boundary is. Distances at which we will eventually get to see something happen are a natural property from the passive perspective of astronomy where we sit at home with our telescopes and wait for light to reach us. However I think that the question of how much of the universe humanity might be able to travel to or affect in any other way is a much more important one — especially since it is unclear why one would care about whether an event in a galaxy billions of light years away was happening right now (as opposed to a million years ago), whereas it is easy to see why we care about the region we could affect starting now.

I believe that this passive way of defining the boundary has made it more difficult for cosmologists to recognise its pivotal importance to understanding the evolution of the universe and the role intelligent life may play in it. Indeed, to my knowledge, the region within this boundary — corresponding to all parts of the universe that we could causally affect — has not even received an official name. I propose that it should be called *the affectable universe*. As we shall soon see, this name is warranted by its tight connection to the observable universe. And I hope to show that the affectable universe is just as important for understanding the causal structure of the universe as its more famous sibling.

### *The observable universe*

Examining our future light cone led us to the idea of the affectable universe. What about our past light cone? That corresponds to all the events which could causally affect us at our current location in time and space. Equivalently, these are all the events that we can observe from our current vantage point. The further back in time we look, the larger the region of space we can observe. If we look back all the way to events that occurred at the time of the Big Bang, we can observe events happening out to 46.4 billion light years away. As with all these numbers, this is in comoving coordinates, so some care in interpretation is needed. It means that we can see the primordial state (just after the Big Bang) of regions of space which are now 46.4 billion light years away (they were closer when the light began its journey). These regions will now be full of galaxies, but at the time when the light left them, galaxies were yet to form.

There is a subtlety here in that the universe was opaque until it was about 380,000 years old. By this time a billion light years of light's longest possible journey had already been used up. So the furthest we can actually detect photons is slightly closer — 45 billion light years away, a sphere known as the 'surface of last scattering'. However it may still be possible to detect other forms of radiation such as neutrinos and gravity waves from the full 46.4 billion light years away, so we will keep our focus on this more fundamental boundary.

This sphere with a radius of 46.4 billion light years is known to cosmologists as the 'particle horizon' or 'cosmological horizon'. The volume within this sphere is called the *observable universe*, since it contains all parts of space that we can (currently) observe. Its spatial extent corresponds to the base of our past light cone.

As it happens, our long distance observations are mainly of events lying along the very edge of our past light cone: events from which light has travelled directly to our telescopes. Even though we could theoretically know about events within the body of the cone, in practice the information largely gets lost. This is because the information from the light that reached us too early has not been preserved and because there are very few particles travelling at speeds between 1% and 99% of the speed of light which could bring us the information more slowly. We thus know about space at different distances in different eras — we only know about the primordial universe in locations very far from us and about the more recent universe in locations that are closer to us. However, from a theoretical level, all locations within 46.4 billion light years are observable and this volume constitutes the observable universe.

Astute readers will notice that the radius of the affectable universe (16.5 billion light years) and that of the observable universe (46.4 billion light years) sum to the greatest distance light can travel (62.9 billion light years). This is not a coincidence, but a fundamental fact of our universe. One can see from *Figure 4* that it has to be true, for each light ray of the past light cone passes directly through the current time and place, becoming a light ray of the future light cone. The distance each light ray travels is thus divided between the distance travelled before our time (defining the radius of the observable universe) and the distance travelled after our time (defining the radius of the affectable universe).

The observable universe and affectable universe are closely related. Elegantly, their spatial extents correspond to the bottom of the past light cone and the 'top' of the future light cone. They could be said to be time-reversed 'duals' of each other. One is inward focused, linked to the past and to limits on what we could *know*. The other is outward focused, linked to the future and to limits on what we could *do*. They thus have the same relationship to each other as afferent and efferent nerves, sensors and effectors, knowledge and action. And there are further symmetries too: the observable universe is everywhere we can see; the affectable universe is everywhere that can see us.

Note that the affectable universe and the observable universe should not be thought of as different universes. Instead, they are different, but overlapping, parts of our single universe — much like how one's upper body and lower body are part of the same body. Their names can be thought of as short-hand for 'the currently observable part of the universe' and 'the currently affectable part of the universe'.

We shall denote the radius of the observable universe at a given time $t$ as $r_o(t)$. And the radius of the affectable universe as $r_a(t)$. These are connected by the equations:

(1)     $r_o(t) = d_\gamma(t)$

(2)     $r_a(t) = D_\gamma - d_\gamma(t)$

And thus:

(3)     $r_o(t) + r_a(t) = D_\gamma$

This also makes it clear that the observable and affectable universes are defined relative to a time. In this case the present time of 13.8 billion years after the Big Bang.

As time passes, the observable universe will grow. Next year, its radius will grow by about one light year and we will be able to see about 25 more galaxies (technically the primordial gas from which those galaxies will be formed). Meanwhile the radius of the affectable universe will shrink by about one light year and about 3 galaxies will slip forever beyond our causal reach.[10] Why are the numbers of galaxies different in each case? Because even though the radii sum to a fixed length, the corresponding volumes do not — the volume of the observable universe is growing faster because it has more surface area. The reason they shall grow and shrink by almost exactly one light year per year is that we have defined the scale factor so it is currently 1. In general they will grow and shrink by $1/a(t)$ light years per year.

As we have seen, our affectable universe (radius 16.5 billion light years) is smaller than our observable universe (radius 46.4 billion light years), meaning we can see things we will never be able to affect.[11] However, this was not always the case. In the early universe, there had been little time for light to travel to one's location, but lots of time remaining for light to travel past many galaxies before the expansion makes the gulfs impassable. This means that from an early vantage point, there were galaxies one could reach which were not yet observable. The point where observable and affectable universes (the top and bottom of the light cones) were equal in size was when the universe was about 4.1 billion years old.

*The eventually observable universe*

What are things like in the distant future, as time approaches infinity? The future light cone gets narrower and narrower (in comoving coordinates), approaching a ray going straight up. The past light cone gets taller and wider, approaching the shape in *Figure 5*, which we could call our *eventual past light cone*. While the radius of the affectable universe approaches zero, the radius of the observable universe approaches the greatest distance light can travel — 62.9 billion light years — and the number of observable galaxies (or locations where those galaxies will form…) more than doubles from its current number. This sphere of radius 62.9 billion light years is aptly known to cosmologists as the 'future visibility limit'. The region within it also deserves a name. I propose we call it the *eventually observable universe*.[12] In some sense this is a more fundamental scale than either the (currently) observable universe or the affectable universe, as its size does not change over time, but rather reflects this fundamental length scale of our universe.

---

[10] These are average numbers of galaxies per year, but they will actually occur in clumps as a gravitationally bound group or cluster becomes observable or ceases to be affectable.

[11] All galaxies with redshift greater than 1.8 are outside the affectable universe.

[12] For symmetry, one could define the *originally affectable universe*, but this is less interesting given that the relevant time has already passed, and it just gives us exactly the same radius anyway.

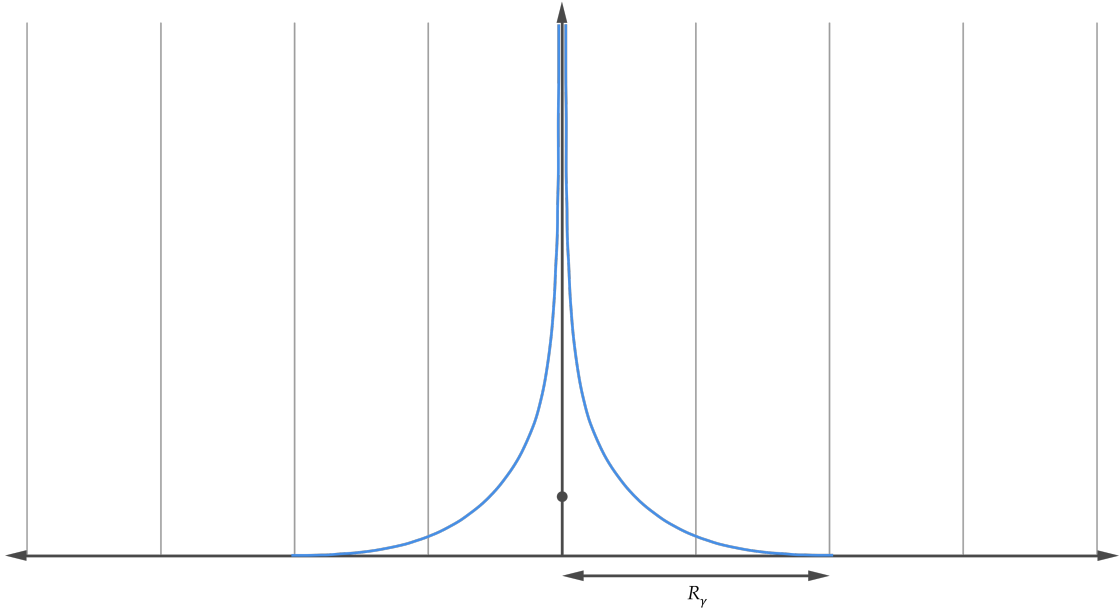

*Figure 5.* The limit of past light cones as time approaches infinity. It's base is the eventually observable universe.

We shall denote the radius of the eventually observable universe $r_{eo}$ and note that:

(4) $\quad r_{eo} = D_\gamma = r_o(t) + r_a(t)$

We have seen how the spacetime diagrams change for events ranging from the start of time, to the infinite future, and have pointed to parts of them (the tops and bottoms of light cones) as representing the affectable and observable parts of our universe. But we may still want diagrams to show how the size of the affectable universe and observable universe change size as a function of time. Happily, *Figures 3* and *5* double for this purpose. While *Figure 3* shows the future light cone for an event at time zero, it also traces out the exact way that the observable universe grows from nothing to the size of the eventually observable universe. While *Figure 5* shows the past light cone of an event at time infinity, it also traces out the exact way that the affectable universe dwindles over time to nothing. One could thus superimpose these curves onto other spacetime diagrams if such information is required.

Finally, observe how some of the comoving galaxies (grey lines) intersect the eventual past light cone. This shows how even if a galaxy can be observed from here, it cannot be observed at all times of its evolution. Even if we waited with our telescopes into the indefinite future, we could only see a fixed initial segment of the galaxy's history, a segment that is longer for closer galaxies.[13] The time from which their future evolution ceases to be observable corresponds exactly to the time at which they leave our affectable universe.

For example, consider the galaxies shown on *Figure 5* at a distance of half the radius of the eventually observable universe. The last time at which our galaxy could have affected that one with a light signal (or vice versa) was 4.1 billion years after the Big Bang, so even if we waited here forever, we would only be able to see the first 4.1 billion years of its history. As our view of it approaches that limiting time, events there would appear to slow down, never quite reaching that time (an effect known as *cosmological time dilation*). The distant galaxy would also become increasingly dim since the photons it emitted over a finite period will arrive here spread out over an infinite period. It would become more redshifted too, as the longer transit times involve more stretching of the light's wavelength during its journey.[14]

---

[13] Only those that are gravitationally bound to us remain observable (and affectable) for ever.

Thus, while the observable universe keeps getting larger as time goes on, there is an important practical countervailing effect. For example, after about 2 trillion years, the scale factor will have grown by about $10^{50}$ from its current value, making the wavelength of all light received from galaxies beyond one's group or cluster grow by this same factor.[15] At this point, even gamma rays would be redshifted enough to make their wavelengths larger than the entire eventually observable universe (and thus larger than any apparatus for detecting them). So while the observable universe is always growing, at some point in the future it may become impossible to practically observe anything outside of one's own group or cluster.

*The ultimately observable universe*

The names 'future visibility limit' and 'eventually observable universe' sound like we have reached the ultimate scale at which we could find out information about distant galaxies. While this is sometimes claimed to be so, it is not the final limit in what knowledge of far-off lands we might one-day acquire.

People making such a claim are implicitly assuming that we will stay in one place, waiting patiently for light to come to our telescopes. But what if we could travel to our nearest galaxy cluster — the Virgo cluster — 50 million light years away? In this case, we would be able to see 50 million light years further in that direction, meaning that eventually around 2.5 billion new galaxies that are invisible from Earth would come into view. This wouldn't increase the total number of galaxies that an observer could come to see — this is still limited by a sphere of radius 62.9 billion light years — but by moving, we would be able to choose where this sphere was centred, and see different galaxies.

---

[14] These are all governed by the ratio between the scale factors at the time light was emitted and observed. If we let $a^* = a(t_{\text{observed}})/a(t_{\text{emitted}})$, then the distant events are slowed by a factor of $a^*$, are dimmer by a factor of $a^*$, and the wavelengths of light are stretched by a factor of $a^*$. So events that we see at a redshift of $z$ are time dilated by a factor of $1+z$. Events that we observe at the (current) radius of our affectable universe appear to happen about three times slower than usual.

This connection between redshift and the scale factor would be even tighter if redshift had been defined as the factor by which the wavelength has been stretched (i.e. with redshift of 1 representing unshifted light). Since redshift is a multiplicative effect, not an additive effect, this definition would also have been more natural and more convenient.

[15] (Krauss & Starkman 2000, p 23).

The ultimate limits of this can be seen in *Figure 6*. The central part of the diagram is the same as *Figure 4*, showing our (current) past and future light cones. On the right hand side, I've shown a light ray from the time of the Big Bang coming towards us, which asymptotes to the edge of our affectable universe. If the light ray came from any closer, then we could (just barely) interact with it, shining our own ray of light towards it and having them meet in the distant future. It would thus be theoretically possible that we might observe it, if we could send a spacecraft at close to the speed of light to a location near the edge of the affectable universe and then look ahead from there with its telescopes.[16] Of course we don't know how fast we will eventually be able to travel through space, nor how far. But since we know that this is bounded by the speed of light, the plan above is compatible with known physics, whereas seeing any further than that appears to be physically impossible.

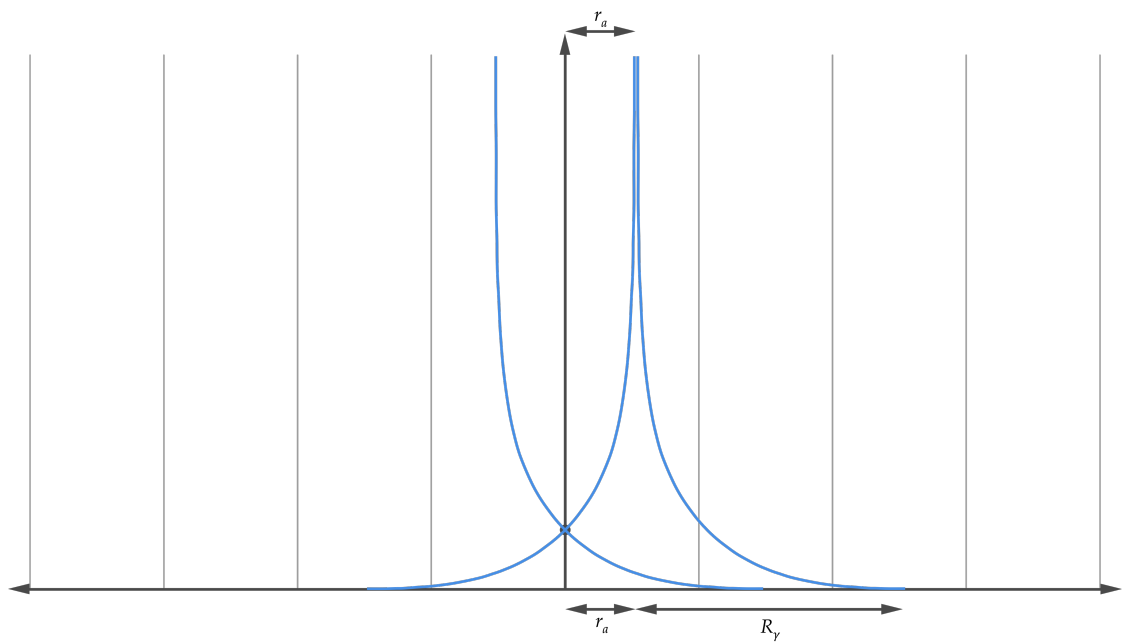

*Figure 6.* A spacetime diagram describing the ultimately observable universe.

I have never seen this immense part of the universe discussed anywhere else, but it certainly deserves a name. I propose calling everything within this radius the *ultimately observable universe*. We can denote this radius $r_{uo}(t)$ and note that:

(5) $\quad r_{uo}(t) = r_a(t) + D_\gamma$

This means its radius is equal to the radius of the affectable universe plus the maximum distance light can travel (for a total of about 79.4 billion light years). Everything beyond this sphere is completely unobservable by us, even in principle. It is this sphere (and not the smaller ones) which may be relevant from the perspective of the philosophy of science, where the question of whether something is *in principle* observable arises. The impossibility of observing anything beyond this sphere has some implications for the testability of physical theories which make claims about what is out there.[17] As mentioned above, I should stress that no single observer could ever see so much, but our descendants could in theory see any part of it they chose, and 'together' their overlapping views might cover it all.

---

[16] This technique would also enable us to see a slightly longer stretch of the histories of galaxies in that direction which were already observable from Earth. For example, the rightmost light ray now intersects the previously discussed galaxy's world line at a time after our present era, instead of about 10 billion years before it.

We are now in a position to show all of these regions of our universe in a single diagram.

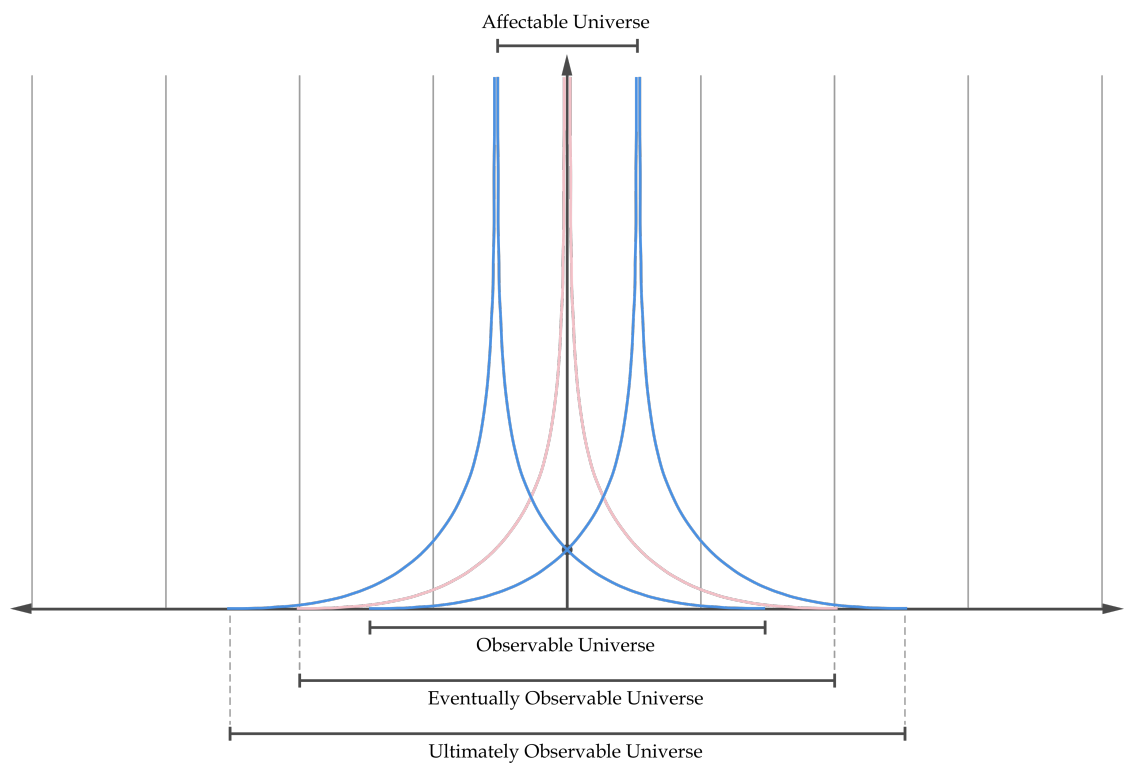

*Figure 7.* The principal causal regions of our universe. The affectable universe is the top of our future light cone. The observable universe is the bottom of our past light cone. The eventually observable universe is the bottom of our eventual past light cone (pale red). The ultimately observable universe is the union of the eventually observable universes centred on each location in our affectable universe.

This diagram suggests an analogy for understanding the different kinds of observable universe as views from the tops of different mountains. We have currently climbed a substantial peak, from which we can see 46.4 billion light years in any direction (the observable universe). By the end of the universe we will have climbed a tall mountain, from which we can see 62.9 billion light years (the eventually observable universe). But there are other mountains nearby that are equally tall. By the end of the universe, we can go to any mountain top within 16.5 billion light years (the affectable universe), and look at the view from its top. While we could still only see 62.9 billion light years from any mountain, anywhere within 79.4 billion light years (the ultimately observable universe) can be seen by travelling to the right mountain and gazing into the distance.

---

[17] I do not mean to say that theories positing events beyond this limit are unscientific, just that this is the relevant distance that such an argument would want to refer to.

*Other boundaries*

In *Figure 6*, we showed the light ray that only just 'touches' our future light cone. In *Figure 8*, we add to this a light ray that only just touches our past line cone. All events to the right of the first of these light rays have future light cones that do not intersect our own. Their futures are causally separated from our own. They cannot interact with us. Correspondingly, all events to the right of the second of these light rays have past light cones that do not intersect our past light cone. Their pasts are causally separated from our own. No event could have affected both us and them.

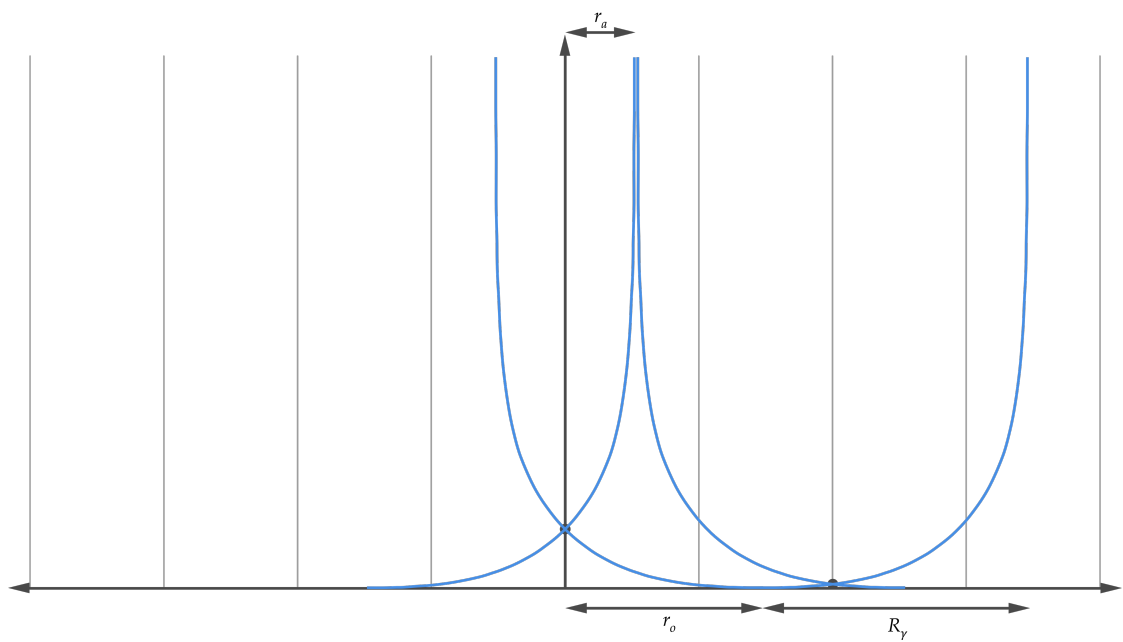

*Figure 8.* Light rays that just miss intersecting our future and our past. These describe the future and past light cones for the closest event that is completely causally disconnected from us (showed by the dot where the rays cross).

The point where these curves cross is the closest event which has no overlap of either light cone. We could call such an event *completely causally separated* from us. *Figure 2* showed how this cannot happen in the simple spacetime of special relativity — pairs of events there always share both a past and a future. In contrast, our universe becomes causally fragmented on a large scale. While there is a chain of causally connected locations connecting any two locations, the beginning and end location in the chain may well share no future or past.

The distance to the closest such event (where the curves cross) is exactly equal to $D_\gamma$ and its time is 640 million years after the Big Bang. This is the time at which the observable universe and affectable universe have the opposite values to those they have today. In some sense it is the 'dual' of our time. While we could eventually observe the initial state of the location where that event would happen (even without leaving Earth), even travelling towards it at the speed of light would only let us see its evolution up to just short of the 640 million year mark.

All events beyond both these lines share no future or past with us. Once we reach a radius of $r_o(t) + D_\gamma$, then every event (no matter at which time it occurred) shares no future or past with us.[18] This forms another causal boundary, but one that seems less important than the others and that we shall thus leave unnamed. Once we reach a radius of $2D_\gamma$, every event (no matter at which time it occurred) shares no future or past with any event that occurred here (no matter at which time that occurred). Thus $2D_\gamma$ forms an outermost causal boundary (which we shall also leave unnamed).

There is a causal boundary much smaller than any of these which also warrants some attention. A radius of $r_a(t)/2$ marks the furthest distance that we could reach and then return from (at the speed of light). Thus, if we wanted to send spacecraft to distant galaxies and have them report their findings back to a central repository, this is the farthest they could theoretically reach before they would have to make their final report. While they could in theory go twice as far as this — exploring the entire affectable universe — the price for this would be eternal isolation from the Earth. As they passed the halfway mark, they could never make it all the way back home (and nor could any signal they sent). This radius of $r_a(t)/2$ also describes the largest volume that is completely causally connected, in the sense that any point within it can (currently) affect any other.

In discussions of the size of our universe, the *Hubble volume* is often mentioned. This is a sphere whose radius is equal to $c/H(t)$ where $H(t)$ is the Hubble parameter at a given time. This is the sphere beyond which comoving objects (such as galaxies) have a proper velocity greater than the speed of light, in a direction directly away from us. (Proper velocities that exceed the speed of light don't involve an object travelling *through* space faster than light — but because the space between us is itself expanding, the objects are becoming further away from us by more than a light year per year.)

There is substantial confusion about how to interpret the Hubble volume, with widespread erroneous claims that we could never affect (or see) galaxies beyond this limit.[19] In fact, this radius is about 14.4 billion light years and is *smaller* than the affectable universe. We can thus affect some of the galaxies beyond the Hubble volume. This may at first sound impossible, since these galaxies are receding at a speed greater than that of light. But the explanation lies in the fact that proper velocities of the things we send after them (including light) also grow as they get further away from us (for the space between us expands) so these can exceed the speed of light too, and sometimes catch up. This will even happen in practice, for if you have ever shone a light into the sky, some of the photons released will eventually reach the edge of the affectable universe, and thus beyond the Hubble volume. By symmetry, some light from galaxies beyond the Hubble volume will eventually reach the Earth.

---

[18] This relies on the fact that for us $r_o > r_a$. The general form of this distance (where all past and future light cones at that distance fail to intersect ours) is $\max(r_o, r_a) + D_\gamma$.

[19] See Davis and Lineweaver (2003) for a list of examples and excellent commentary. They also point out that the term 'Hubble Volume' is sometimes used to refer to the observable universe, which adds to the confusion.

The Hubble volume, correctly understood, is far less fundamental for the causal structure of our universe than the other boundaries discussed here. The only interesting causal property of the Hubble volume that I know of is that in the limit of infinite time, it converges to the size of the affectable universe and thus it eventually comes to share the affectable universe's properties — however this doesn't provide any additional value if we were already familiar with the affectable universe. It is mentioned here just for completeness, and to help dispel lingering confusions.

**Other measures for time and space**

The spacetime diagrams so far have used comoving distance — factoring out the expansion of space. For comparison, *Figure 9* is a version of *Figure 4* using proper distance, where this expansion is not factored out (the time scale has been stretched a little to show more detail).

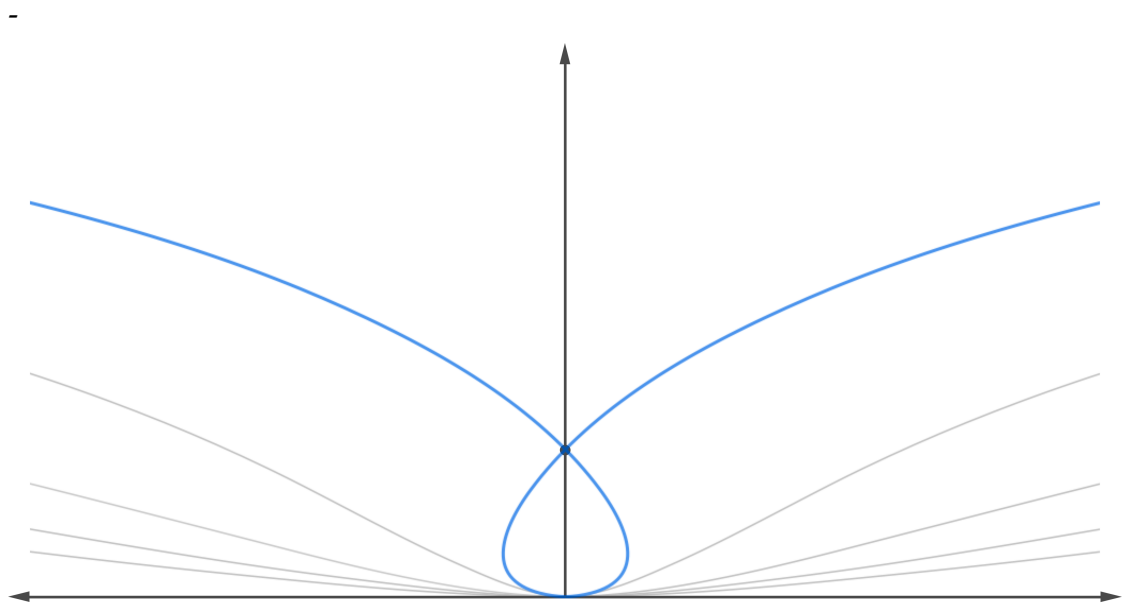

*Figure 9.* A version of Figure 4 using proper distance instead of comoving distance.

We can thus see that the curved nature of the light rays is not just a product of using comoving distances, but is an effect of expanding space. When using proper distance, rather than reaching a fixed size, the future light cone expands more and more quickly. This shows how light released now can travel an unlimited proper distance, but it obscures the fact that we can only reach a finite number of galaxies, since they can recede even faster.

The past light cone is onion shaped, reflecting how everything was extremely close together at the Big Bang. Surprisingly, the light that reaches us now from the right, was initially moving away from us in terms of proper distance, because the space between us and it was expanding too quickly. As the light passed more and more of that expanding space, it eventually reached a point where it could make progress against the expanding background.

The diagram also shows how the comoving galaxies spread out rapidly through space on paths reflecting the changing rate of expansion of space over time (the rate of expansion was initially slowing, but started to increase again around 7.5 billion years after the Big Bang). Since all of these galaxies shown were outside the affectable universe, none will cross the future light cone. If we could zoom in enough, we would see that the world lines of the closest galaxies shown here began within our past light cone.

Note that because the world lines of these galaxies are not vertical, light cones drawn around them would not have the same shape as light cones drawn around us. This makes spacetime diagrams using proper space much more difficult to use.

Cosmologists also use a third kind of spacetime diagram, where the horizontal axis is in comoving distance, but the vertical axis is distorted in such a way that light rays take straight lines. More precisely, it plots comoving distance against *conformal time*, τ. This is defined using the same function that governs how far light can travel:

(6)    $\tau = d_\gamma(t)/c$

The infinite span of future time is thus compressed into a finite range. In the same way that light can travel a maximum comoving distance of 62.9 billion light years, conformal time has a maximum of 62.9 billion conformal years. This means that the top of a conformal spacetime diagram corresponds to a point in the infinite future.

The payoff for this strange temporal distortion lies in what it does to spacetime diagrams. Aspects of the causal structure of spacetime that were previously obscure become much clearer. We can see how this works in *Figure 10*, which represents the same thing *Figure 8*, but in conformal time. Like *Figure* 8, it shows our time and location near the centre, with our light cones spreading out, and also the nearest event in space whose light cones do not overlap ours. (For reference I have also added a some pale red lines to the diagram so that it also shows the largest past and future light cones centred on us.)

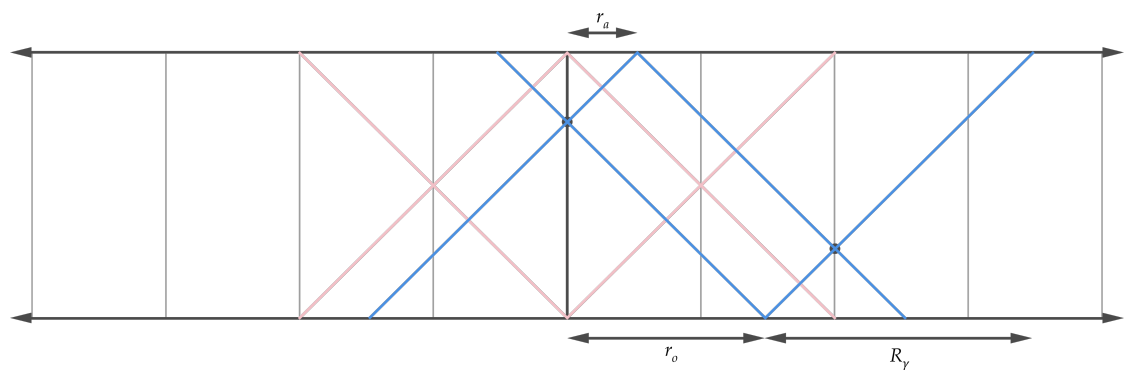

*Figure 10.* A version of *Figure 8* using conformal time. I have also added the largest past and future light cones in pale red for reference.

We can see that the radii of the different spheres around us has been left unchanged, but the compression of time has made all the lines straight and the structure easier to understand. Where the affectable universe corresponded to the limiting size of the future light cone as time approached infinity (infinitely far up the diagram), the diagram now includes this temporal point (where $t = \infty$ and $\tau = D_\gamma / c$) so now the affectable universe is simply the spatial slice of the future light cone at the top of the diagram.

We can also see that, in terms of conformal time, we are quite late in the universe, being about three quarters of the way to the top. This corresponds to us being at a time where light has travelled about three quarters of the eventual (comoving) distance it will travel. The closest event to us which we cannot interact with has a conformal time that is exactly as close to the bottom as we are to the top. This supports my earlier suggestion that its time is the 'dual' of our current time.

In comoving coordinates and conformal time, the spacetime structure of our universe looks very similar to the simple structure of spacetime from Special Relativity (see *Figure 1*), just with a finite beginning and finite end. For some purposes this is indeed a helpful simplification. However, since neither space nor time on the diagram no longer have their intuitive meanings, care must be taken if using this analogy to form intuitions or derive conclusions.

Finally, we can draw a new version of *Figure 7* using conformal time, showing very clearly the major parts of the universe we have described.

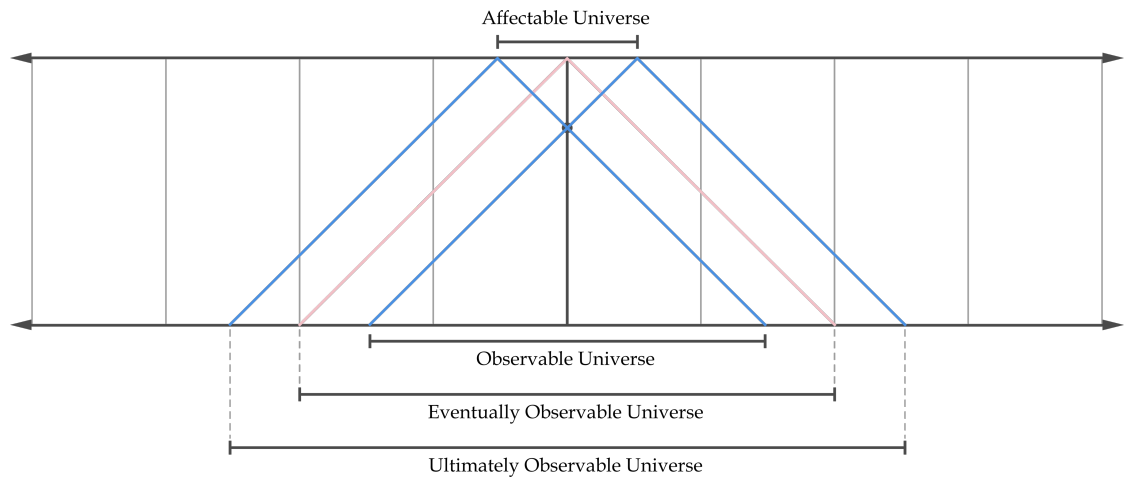

*Figure 11.* The principal regions of our universe, shown with conformal time.

## PART TWO — APPLICATIONS TO INTERSTELLAR CIVILISATIONS

### Travel below the speed of light

The analysis above has covered how far signals could travel at the speed of light. It would also be useful to extend these concepts to speeds below *c*. Given certain assumptions this is relatively straightforward. If something can travel at a constant fraction of light speed, *fc*, then at any time, it will have reached *f* times as far as light. This also implies that the maximum (comoving) distance it could reach is *f* times as far as light can reach. Thus, the 'travel cones' of things moving at a constant fraction of *c* are just horizontally compressed versions of the light cones we have seen earlier, where the degree of compression is linear with the speed. This is true regardless of which kind of diagram we are using.

We can therefore create analogues of the earlier concepts. For example, if *Fc* is the greatest achievable constant speed at which spacecraft could travel long distances, then the reachable radius with such spacecraft starting at time *t* is $Fr_a(t)$, and the furthest we could see is $Fr_a(t) + r_{eo}$.

It is important to note that it is not trivial to travel at a constant fraction of *c* (even a small one). The expansion of space reduces the momentum of all particles travelling through it. For light, its *speed* cannot be reduced so the momentum reduction takes the form of redshift. For particles with mass, it takes the form of reducing their speed (their proper speed, relative to things they are passing). Thus a spacecraft launched at some fraction of *c* will slow down below that fraction unless something is done to keep up its speed. This could take the form of continuous thrust or by breaking the journey into many shorter trips and reaccelerating up to top speed at the start of each of them. If there were enough of these shorter trips, then the changes in the speed would make very little difference to the calculations and we could treat it as travelling at its average speed. As galaxies are typically millions of light years apart, the periods of coasting through intergalactic space are likely to be long compared to the periods of acceleration, deceleration, and preparing for the next journey. The average speed of travel over the long haul is thus likely to be close to the top speed.

It may also be possible to make use of the reduction in momentum. A major challenge in sending a spacecraft to a distant location is how to decelerate when arriving. While rockets can do this, the fuel needed to do so would greatly increase the size of the craft and thus the energy needed to launch it. However, over long distances the deceleration due to the expansion of space could help with this challenge (at the high cost of reducing the reachable distance by as much as 50%).

Our earlier analysis assumed that the distribution of matter in the universe is relatively homogenous — an assumption that holds at the distance scales that light can reach. Cosmologists often use a figure of about 300 million light years to represent the scale at which things can start to be treated as homogenous.[20] One

---

[20] This figure is determined by fixing a radius, looking at spheres of that radius around randomly chosen locations in space, and comparing the average amount of mass in each sphere to the standard deviation. Above a radius of 300 million light years, the standard

would need to travel at about 2% of *c* in order to reach this scale, so below this speed, the results in this section will be notably sensitive to the local distribution of galaxies around one's starting point. The escape velocity of the Milky Way from Earth's location is about 0.2% of *c*, so below that speed, a civilisation may be restricted to its starting galaxy, in which case very little of this analysis would apply.[21]

**Timings**

*Time of departure*

How soon would our civilisation need to set out towards the far reaches of space in order to reach most of what is currently reachable? We can determine this from our knowledge of the current size of the affectable universe and how it will change over time. Is there cause to rush?

Not really. Each year the affectable universe will only shrink in volume by about one part in 5 billion. Even if we waited a million years, it would only diminish by one part in 5 thousand. It would take 50 million years, before its volume would shrink by 1% (see *Table 1*).

In absolute terms, these would be losses on a scale beyond normal comprehension: the loss of 3 entire galaxies for every year of delay — thousands of stars every second.[22] But for most purposes it is the relative scale that matters. For example, if through a year of delay we could improve spacecraft speeds by even 1 part in 5 billion it would be worth it. If by going more slowly we could take on less risk, and improve our chance of making it to the launch date by just 1 part in 5 billion that would also be worth it. As would a 1 part in 5 billion improvement in the quality of what our civilisation would accomplish in each galaxy it reaches. So in the end, it may not matter much (proportionally speaking) if we waited millions or tens of millions of years before setting out. If this would perceptibly improve our speeds of travel, our chance of success, or the quality of what was achieved, it would be worth the delay.

However, by the time we get to delays on the order of a billion years, a significant amount of the universe would be lost to us. By roughly 150 billion years, we would be unable to leave the Local Group, and thus would lose everything that could be

---

deviation becomes smaller than the average. This is still a fairly weak level of homogeneity. For reference, this sphere is about the same size as our supercluster, Laniakea, and there are voids which are larger than this. For contrast spheres with 5 times this radius (corresponding to the distance we could reach at 10% *c*) would contain more than a hundred superclusters and would be much more homogenous.

[21] There are ways one could try to get around this, such as by progressively moving further and further out from the centre of the galaxy, so that the escape velocity from the new location is lower, and by utilising stars that float between the galaxies as way points.

[22] Bostrom (2003) expresses this vividly (though the details of his argument differ).

lost, leaving us with only 55 galaxies out of the 20 billion which are currently affectable.

| Delay | Galaxies lost |
|---|---|
| 1 million years | ~ 0.02% |
| 10 million years | ~ 0.2% |
| 100 million years | ~ 2% |
| 1 billion years | ~ 20% |
| 10 billion years | ~ 80% |
| 150 billion years | ~ 99.9999997% |

*Table 1.* Proportion of the reachable universe lost due to delay.

The results are very similar for travel at any speeds between 2% and 100% of *c*. While the absolute effects of delay are lower with a smaller maximum speed, the proportional effects are largely unchanged. The main exception is the final row of this table, but even then it doesn't change much. For the Local Group to become isolated to spacecraft travelling at *fc*, the scale factor only needs to be *f* times what is required for it to be completely isolated. For example, the Local Group becomes isolated for spacecraft travelling at 2% *c* when the scale factor reaches 2% of what it will be in 150 billion years — this will happen about 80 billion years from now.

*Time of arrival*

While it could take an arbitrarily long time to arrive at all reachable galaxies, most of them would be reached while the universe is roughly like it is now. Travelling at *c* (or a constant fraction of *c*) we can travel to 50% of the reachable galaxies within 30 billion years, 90% within 60 billion years, and 99% within 100 billion years (less than ten times the present age of the universe). We will thus arrive almost everywhere we can reach at a time when stars are still burning — though many of the existing ones will have burnt out and there may be a much higher proportion of red dwarfs. New stars will still be getting created (until 1 to 100 trillion years). And many currently existing stars will still be burning (the smallest red dwarfs are estimated to burn for trillions of years).

*Timescale of civilisation*

A civilisation could only reach this ultimate scale if it could endure in some form for billions of years. This may be somewhat easier than it sounds, because it would be able to have independent settlements across many planets, stars, galaxies, groups, and clusters. This means it would be much more robust to disasters, in much the same way that a species is more robust and long lived than each organism of that type.

Extinction isn't the only risk though. Even if it could reach new worlds at a faster rate than the existing worlds die, it would have to face the possibility of its values drifting so far that its original aims would fail to be achieved, with something very different taking their place. It is unclear whether this would be generally be for the best or for the worst, but it is certainly something that civilisations would take account of and may strive to prevent. It is also a change so great that we might be tempted to say the original civilisation did not *survive*, but was replaced by one or more new civilisations.

There are other timescale considerations as well. To be able to move between galactic groups at all, spacecraft would need to survive in intergalactic space for millions of years while retaining enough capability to decelerate into their destination galaxy and create a new outpost of civilisation there. There are several strategies for having a spacecraft survive over such lengths of time, including resistance to damage, self-repair, and redundancy (within each spacecraft or by sending multiple spacecraft).

If very high speeds are possible, then relativistic time dilation could reduce the subjective journey time by a large factor. However this would bring its own challenges. It requires a very large amount of energy to launch the spacecraft (the kinetic energy required is proportional to the time dilation factor), it greatly magnifies the danger of hitting interstellar dust (increasing the energy of impacts in proportion to the time dilation factor), and it greatly magnifies the problem of how to decelerate at the journey's end.

In short, survival of spacecraft over such large timespans would require very impressive feats of engineering and it would be particularly difficult for flesh-and-blood humans to make such journeys.

*Eras of the universe*

We have seen that after about 150 billion years, the Local Group will be isolated, with all other galaxies having left its affectable universe. This will happen at a similar time for other groups and clusters.[23] The history of the universe can thus be roughly divided at that point into two important eras.[24] I propose that we call these: *the era of connection* and *the era of isolation*. At typical times within the era of connection there are billions of affectable galaxies, whereas by *the era of isolation* any galaxies that are not gravitationally bound to each other can never again affect one another, shrinking the number of affectable galaxies by a factor of a billion.

Ambitious civilisations surviving over such timespans would face different challenges in each era. The era of connection would involve moving quickly to reach distant galaxies before they become inaccessible. It would also involve solving the challenging problems of coordinating these different outposts of civilisation, keeping their goals in harmony with each other during the time when problems of ideology or war in one location could spread to others, increasing the risk of correlated disaster through a large region of space. While light speed communication is slow on these grand scales, it may still play an important role, since it is still possible to have hundreds of round-trip communications between nearby galactic groups before they become isolated. Finally, this era would also involve setting the outposts off in a good societal direction before they become isolated, so that they continue advancing whatever grand ends the civilisation was pursuing.

---

[23] It will take the longest with very small groups which are close to each other, but just far enough away that they don't remain bound. Busha et al (2003, p 718) show that even stars that are not bound to a galaxy must be isolated within 336 billion years and unbound particles become isolated within 1,060 billion years.

[24] Busha et al (2003) also stress this cosmic division of time.

The era of isolation would involve each outpost having a truly vast amount of time at its disposal. It would be able to patiently work to pursue its ends, with its fundamental deadline set by much slower physical processes such as the trillions of years before the stars go dark or the truly astronomical times before matter decays or black holes evaporate. As we will see in the section on computation, this is the time period in which most computation could be done and in which the most lives could be lived. The isolation also means that a disaster in one region would no longer be able to spread to the others, making things much more robust, and potentially allowing each outpost much more liberty in how it is run.

The era of isolation also imposes a maximum physical scale on any projects that could be undertaken (and on any causal relationships whatsoever). Consider the Local Group in 150 billion years. At this point, it will consist of the merged Milky Way and Andromeda galaxies surrounded by a tightening cloud of dwarf galaxies (which will have all merged together by about 450 billion years). Over much longer timescales, smaller stars will be ejected from this merged galaxy, gaining enough kinetic energy from close encounters with other stellar remnants to reach the Local Group's escape velocity and drift away. Surrounding the merged galaxy is an immense void — nearby galaxies long since departing via cosmic expansion.

Continuing expansion puts a limit on how far an object could go into this void and ever be able to return. This limit is the radius of the affectable universe at that time. We have previously discussed this limit in terms of comoving coordinates and seen that it shrinks towards zero as time goes to infinity, making other galactic groups and clusters unreachable. However, here we are more interested in its size in proper distance. This asymptotes to about 17.3 billion light years in radius. The further something is from a bound galactic group, the faster it needs to travel to be able to outrace the expansion of the intervening space and eventually return. The 17.3 billion light year sphere marks the point of no return — a distance beyond which even the speed of light would be insufficient to come back.[25] All isolated matter (be it a group, a cluster, or even a lone star or particle) will have such a sphere around it of this same size. If anything drifts beyond this, it can never return.

---

[25] There is thus a rough analogy between this sphere and the event horizon of a black hole, which is why this sphere that bounds the affectable universe is often called the universe's event horizon. Though note that in this case, objects are being 'sucked' *out* towards it, so it is something like an inside-out black hole.

This puts a (very large) upper bound the kinds of structures that can survive in the long term. The main relevance of this bound is not in its particular size, but the mere fact that it is finite. Some approaches to achieving indefinite survival of life and an indefinitely large amount of computation over the long-term future,[26] require structures whose size grows without bound as time goes on. This size limit would seem to prevent such a strategy.

**Information**

The spatial region from which we can currently gain information is the observable universe. The greatest spatial region from which a single entity can gain information is the eventually observable universe around a point (such as Earth). The union of all parts of space we could ever find out about is the ultimately observable universe (though any individual can only find out about a sub-region of this). If we wish to take into account the time periods that are observable at each location, we get the past light cones corresponding to each of the above (the union of a set of past light cones in the final case).

**Resources**

*Matter & Energy*

The matter distribution of the universe gives rise to a small number of natural sizes that a technological civilisation might reach: that of a planet, solar system, or galaxy. These natural stopping points exist because it appears to be substantially easier to progress towards perfection of one level than to move up to the next. The famous Kardashev scale uses this observation to classify technological civilisations into three levels. Kardashev was interested in the question of how much power (energy per unit time) an ambitious civilisation might be able to access, which he took to be the maximum amount of starlight that could be harnessed at each of these size scales. But the same classification makes sense for other questions too, such as the amount of matter (or matter of a given type) accessible at each level.

Many people have tried to extend the Kardashev scale to a fourth level, corresponding to 'the universe'. But to my knowledge, none of these attempts have succeeded, since they lacked the right concept of universe. Given our current understanding, it is physically impossible to settle the entire universe, or even the observable universe. But there *is* a natural level, which is the affectable universe.[27] (Or, at least roughly so. If the maximum speed at which any physically allowable spacecraft can travel to and settle distant galaxies is substantially lower than the speed of light, then that speed would give the right measuring stick, replacing light cones with travel cones, and the affectable universe with the settleable universe.)

---

[26] Such as that of Dyson (1979).

[27] I suggest this development in Appendix G of *The Precipice* (Ord, 2020).

Interestingly, an intergalactic civilisation which reached the size of the affectable universe would continue the pattern of exponential scale-up in accessible power between each of these levels (*Table 2*). One could also extend the scale in the other direction by asking about the starting scale of civilisation. Taking that to be the size of ancient Mesopotamia at the dawn of the written language, we find that this level 0 also fits the same pattern.

| Level | Civilisation Size | Scale-up | Power |
|---|---|---|---|
| K0 | Initial | | $\approx 10^8$ W |
| K1 | Planetary | × 1 billion | $2 \times 10^{17}$ W |
| K2 | Stellar | × 1 billion | $4 \times 10^{26}$ W |
| K3 | Galactic | × 100 billion | $4 \times 10^{37}$ W |
| K4 | Ultimate | × 1 billion | $4 \times 10^{46}$ W |

*Table 2.* The extended Kardashev scale, with a level K0 set by the starting level of civilisation and level K4 set by the size of the affectable universe.[28]

These particular numbers are based on our own situation: the first civilisation on Earth, the amount of sunlight Earth intercepts, the luminosity of the Sun, the luminosity of the Milky Way, and the size of the affectable universe at the current cosmic time. Each of these may be somewhat different for a civilisation on a different planet, around a different type of star, in a different sized galaxy, at a different cosmic time. However, the levels should usually still be separated by several orders of magnitude — at least for civilisations that arise before the era of isolation.

One could logarithmically interpolate between these levels to find where transitional civilisations lie. For example, humanity currently controls about 12 trillion Watts of power. This is about 100,000 times more than a minimal civilisation but 10,000 times less than the full capacity of our planet. This would place us at level K0.55 — more than half way to K1 and an eighth of the way to K4.

But why concern ourselves only with the instantaneous amount of energy per unit time? What are the total amounts of matter and energy that could be harnessed on this largest scale?

The amount of matter that a civilisation could affect is bounded by the affectable universe. The amount it could send spacecraft to is bounded by a smaller travel sphere, whose radius is $F$ times as big if its top speed of travel over this distance is $Fc$. Since the volume increases as the cube of the distance, this can make a large difference. A top speed of 10% $c$ would mean only a thousandth as much reachable matter as for a top speed of $c$.

The affectable universe contains about 20 billion galaxies with a total of between $10^{21}$ and $10^{23}$ stars (whose average mass is half that of the Sun).[29] This region contains a very large amount of energy, that comes in many different forms. The total energy of the affectable universe is around $10^{53}$ kg (or $10^{70}$ J).[30] Of this energy, about 69% is dark energy — the unknown form of energy that permeates space and is responsible for the accelerating expansion. On the ΛCDM model (which this paper takes as a starting assumption) this dark energy has the very strange property of not diluting as space expands — it maintains a constant density of about $7 \times 10^{-30}$ g/cm$^3$. Thus when the matter becomes more sparse due to cosmic expansion, dark energy's share of the total energy density will approach 100%. The following diagram therefore treats dark energy separately and concentrates on the relative shares of the other forms of energy — most of which is matter. While the total density of matter decreases towards zero as space expands, the relative shares won't change as much, and this should remain a good guide over very long periods.

---

[28] Adapted from Ord (2020). There I put K0 as 'Minimal' but I now think 'Initial' is more accurate.

---

[29] We don't know precisely how many galaxies there are. Using a recent estimate of 0.0009 galaxies per cubic mega-light-year at the present moment (Conselice et al., 2016), I calculate around 17.5 billion in the affectable universe (which I round to 20 billion to reflect the uncertainty) and 400 billion galaxies in the observable universe. Conselice et al arrive at a different number for the latter because they were integrating over the past light cone (i.e. counting galaxies in earlier periods of the universe the farther out they look) and the number changes over time as galaxies merge together. Galaxies come in a vast range of sizes, which adds to the uncertainty. We may find that there are many more small and faint galaxies than we had anticipated, greatly increasing the number of galaxies in the observable universe, but simultaneously making the average galaxy less impressive.

[30] We measure the energy in kg via the mass to energy equivalence of $E = mc^2$.

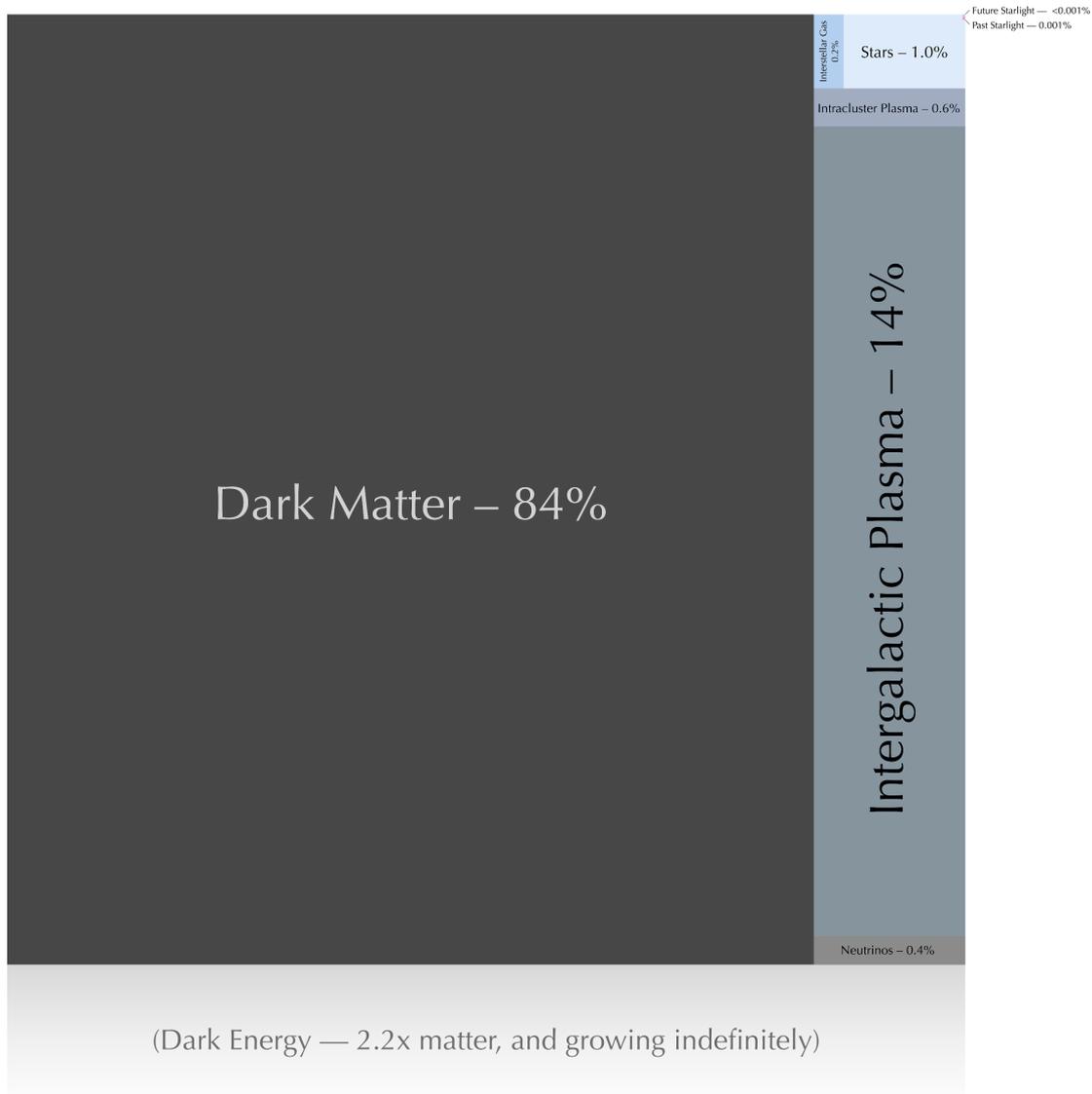

*Figure 12.* The composition of the mass-energy of the universe. Note how little is in the form of stars, and how only a tiny fraction is in the light released by stars. [31]

As mentioned earlier, matter on these cosmological distance scales is distributed roughly homogenously and astrophysicists also suspect that each large-enough region will contain roughly the same proportions of all common naturally occurring objects, such as elements, compounds, stars, black holes and galaxies.

This provides upper bounds on the amount of matter and energy a civilisation could ever access. However, it is very unclear whether a civilisation could really come close to these upper bounds.[32]

---

[31] The main breakdown is the latest data from Planck Collaboration (2016). The breakdown of ordinary matter is from Fukugita and Peebles (2004).

[32] Tegmark (2017) contains an interesting exploration of how much energy might be harnessable, and indeed touches on many of the other themes of this section concerning ultimate limits.

Regarding the matter that could be accessed, note that we currently have very little idea how to usefully interact with dark matter, and that most of the remaining matter is in the intergalactic plasma. This is so thinly distributed through intergalactic space that it may also be impossible to make use of. Together these comprise 98% of all matter, meaning that there is very wide uncertainty in how much matter could be harnessed.

In terms of the available free energy, note that less than 0.1% of the mass of stars will be radiated as starlight (less than 0.001% of the total mass). So even if a civilisation could capture all of that, there is the potential for much more energy if some of the mass-energy of the stellar remnants or other components of the cosmos could be liberated. For example, dropping matter into black holes via their accretion disks converts 5.7% of its rest mass into light, which far exceeds the efficiency of the thermonuclear reactions which power stars, and might be a much more efficient way to gain access to the mass-energy stars contain (though it clearly comes with its own engineering obstacles). If through this or some other technique any reasonable fraction of the mass energy of ordinary matter can be harnessed, then the light from all the stars would make up only a tiny fraction of their usable energy, and the starlight lost before arriving in a galaxy and setting up solar collectors would be insignificant. (Note that if we can extract useable energy from the dark energy, it would dwarf everything else since it will grow without bound as the universe expands.)

A related question is not how much matter (or energy) could be reached, but what is the greatest amount a civilisation could secure in one gravitationally bound location, so that it could be used for some unified long-lasting project during the time of isolation. A lower bound can be formed by finding the largest pre-existing bound structures that a civilisation could travel to. For instance, the Virgo cluster has around 1,000 galaxies and other, larger, clusters could be detected and visited. A generous upper bound would be all of the matter within a sphere of half the radius of the affectable universe (since it is impossible to reach matter beyond this point and still return). This is around 2 billion galaxies. A tighter upper bound could be created by taking into account how much mass would be wasted in an optimally efficient method of gathering matter together.

*Computation*

While less general than matter or energy, it might be interesting to ask how much computation a spacefaring civilisation could potentially perform. In particular, we could ask about the largest serial computation a civilisation can perform, the largest parallel computation it can perform, or about the total amount of computation it can perform.

The total amount of computation is closely related to the amount of matter and energy the civilisation could reach (bounded by the affectable universe). Unlike a parallel computation, there is no requirement here that the computational threads come back together to provide a single result. So the best strategy would appear to be to spread out as far as possible and to do the computation locally in each cluster or group, using the local resources in an optimally efficient way.

Surprisingly, it appears that this efficient way of maximising the amount of computation performed in each galaxy involves waiting an extremely long time before beginning, and then computing extremely slowly.[33] This is because Landauer's principle tells us the minimal energy needed to erase a bit of information is proportional to the temperature and lower temperatures will be available in the future.

The cosmic background radiation currently sets the floor on cold temperatures we can achieve without using energy. It is currently 2.7 degrees Kelvin and is inversely proportional to the scale factor. So for example, in 500 billion years, when the scale factor reaches 1 trillion, the background would be a trillionth as warm, and we would only need a trillionth as much energy to erase a bit. If this is the limiting condition on the amount of computation, then we could then perform 1 trillion times as much computation for the same amount of energy. In around 1.4 trillion years, the temperature will reach a floor of $10^{-30}$ K and stop declining any further. This residual temperature comes from horizon radiation, which does not decline over time, so this may represent a good time to begin the computation.[34]

As well as waiting a long time before beginning, the civilisation would need to compute extremely slowly, since each time a bit is erased, the machine would warm up and would take a long time to cool back down to its efficient operating temperature. The only deadline on this computation would be the instability of the computing substrate.

It should be noted that that this approach to overcoming the limit imposed by Landauer's principle pushes the limit back so far that other limits may bite significantly earlier, reducing this apparently enormous scale-up of our computing resources. For example, it may be impossible to create physical hardware capable of reliably computing at anywhere near such cold temperatures. Thus this should only be taken as an upper bound.

There are two main strategies a civilisation could apply to perform extremely large parallel computations: (1) do all the computation within a gravitationally bound structure (natural or constructed) or (2) spread the computation over distant galaxies, which all send their finished parts of the computation back to a central point. The first approach has been covered by the earlier discussion on largest bound structures and using the matter in a structure to perform as much computation as possible.

---

[33] See Sandberg et al (2017). This utility of this approach has been questioned by Bennett et al (2019).

[34] This ultimate temperature floor and its implications for long term information processing has been noted by Gott (1996), Barrow & Tipler (1996), and Krauss & Starkman (2000).

The second would involve using galaxies at a distance of up to half the radius of the affectable universe (as the civilisation needs to reach them and then have a signal return). At the present time, this includes around 2 billion galaxies, which is about a million times as many as in a natural cluster. However, despite this increase in resources, if it is true that computation is much more efficient when the universe is colder, then this strategy won't help much compared to just using a single natural cluster. Even the nearest galactic groups would have to send their signals back to the central system before they become isolated (in about 150 billion years). At this point, the background temperature is about 1 in 6,000 of its current value, allowing 6,000 times as much computation as if the energy was used at the current temperature. But in, say 500 billion years, the background would be about 1 billion times colder than that, allowing a billion times as much computation (assuming other limits don't bite first). Thus even a billion galaxies helping in the computation may only do about as much computation as a single galaxy could in 500 billion years' time. So the contribution of all these other galaxies (which have to return their parts of the solution while the universe is relatively warm), might be insignificant compared to what could be computed in a group or cluster when things are colder.

The largest serial computation is covered by the above analysis. It would involve the largest possible cluster of matter (whether natural or formed for this purpose) and using the energy from it to perform a very long, very slow computation in the cold, distant future.

**Alien Life**

So far I have set aside questions of interactions between independently originating forms of life, and have instead considered the upper limits of what could be achievable in an otherwise lifeless universe. There are many ways that alien life could change this picture — especially its role in changing a civilisation's ambitions. For example, a civilisation that might have aimed to bring life to all corners of a barren universe, may instead turn to exploration and stewardship if it finds the universe filled with non-intelligent life. Or, if it finds other intelligent life, it may turn to conversation and exchange of ideas. Properly exploring such possibilities would take us far beyond the scope of this paper. But it is worth noting a few ways that the different causal boundaries we have examined would have direct effects on questions about alien life, or the interactions between independent civilisations.

For example, an understanding of these edges of the universe is necessary to properly articulate questions regarding whether we are alone in the universe. This is often expressed in terms of whether we are alone in the observable universe, but that may not be the most useful question. The main consequence of being alone in the observable universe is that it would be impossible to detect alien life from Earth at this moment. But we may still be able to do so if we wait or travel. Thus even if no alien life ever did or ever will begin within the (current) observable universe, we could still have a future that is influenced by more distant alien life. To get the more intuitive, but stronger, conclusion that humanity couldn't ever interact with alien life, we would have to be alone in the ultimately observable universe.

As we have seen, how far a civilisation can reach depends closely on what time it begins. The results above about being alone in the universe are mainly driven by the possibility of very early civilisations, with correspondingly vast causal reach. But what if we ask about civilisations at our own cosmic time? In order for the current and future actions of an alien civilisation to ever have any effect upon the Earth, that civilisation would have to be within our affectable universe. Even if it was beyond this distance, it may be able to affect our descendants by reaching places that they could also reach. In order for its current and future actions to be able to affect anything we could affect, the civilisation would have to (currently) be within a distance of $2r_a$ of Earth (such that our affectable universes overlap).

We could give more general bounds by also taking the time of origination of the alien civilisation into account. This can be done by replacing the spherical regions above with the relevant spatiotemporal regions. For example, to ever affect the Earth an alien civilisation would not merely have to arise within our eventually observable universe, but within our eventual past light cone. We could also add in constraints about travel at speeds below *c*, replacing some of the light cones in this analysis with analogous 'travel cones'.

When estimating the probability of spacefaring alien civilisation originating in a given region of space at a given time, we benefit from improved understanding of the origination of life on Earth and also from astronomical information about the frequency and characteristics of planets, the lifetimes of stars, and so forth. For most of this information we do not gain that much from having access to extremely distant observations, but for one key parameter we do — whether or not there are any large-scale signs of alien civilisations. For example, if there is a reasonable likelihood that the type of civilisation that would spread through space would also engage in stellar- or galactic-scale engineering projects, then we can get more information by seeking the signatures of such engineering projects in galaxies at all ranges. For this kind of information gathering, the entire observable universe is useful.

Interestingly, if we were to observe a spherically expanding intergalactic civilisation deep in the sky, it would not *appear* spherical.[35] It would instead have an asymmetrical egg-like shape, with the end towards us being more pointed (see *Figure 13*). This distortion doesn't come from cosmic expansion — the appearance is the same as it would be in a flat, non-expanding, spacetime. Instead, it is due to the finite speed of light.

---

[35] Olson (2015, 2016, 2017, 2018) has developed much of the theory of expanding cosmological civilisations and explored its consequences for observational astronomy, including these shapes.

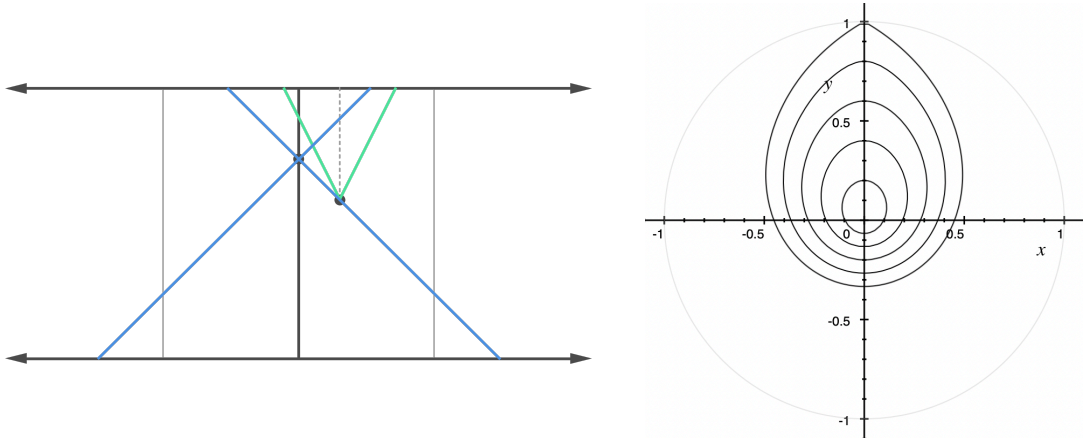

*Figure 13.* LEFT. The conformal spacetime diagram showing the moment when an alien civilisation is first detectable. It's expanding domain is shown in green with a travel cone at 50% the speed of light. Note how at the moment it becomes detectable, it already extends half way to Earth. RIGHT. The changing apparent extent of the alien civilisation. It expands spherically from the origin and Earth is located at the point (0, 1). The civilisation would be invisible until it had actually expanded 50% of the way to Earth, at which time it would be seen as a point at the origin. The black curves are how it would appear when it was actually 60%, 70%, 80%, 90%, and 100% of the way to Earth. At all these points in time its true extent would be a larger circle around the origin. The grey circle is its true extent at the time it reaches the Earth.

The precise curve of one of these shapes can be calculated by considering the set of all points in space where the time taken at the maximal travel speed to get from the centre of expansion to that point plus the time at light speed to get from that point to the observer is equal to the same constant.[36]

Alternatively, one can understand this shape in terms of a conformal spacetime diagram. It is simply the spatial region given by the intersection of two cones — our past light cone and their future travel cone. (Indeed many spatial regions of interest in this field are given simply by the intersections of cones.[37])

Knowing the shape to look for, one could sift through galactic sky survey data to search for irregularities fitting this shapes. For example, a region whose galaxies have a different emissions spectrum, or where there are fewer visible galaxies than expected. If such a region were detected, we could back-calculate the point of origin, the travel speed, and thus the actual current extent of the alien civilisation. This is a promising avenue for SETI, as even if only a small fraction of alien civilisations reached this scale, this may be more than compensated by the fact that we are searching for evidence of alien life in billions of galaxies at once.[38]

---

[36] This can be done with the equation $\sqrt{x^2 + y^2} + F\sqrt{x^2 + (y-1)^2} = k$, where the alien civilisation starts at the origin, travels at speed *Fc*, and distances are normalised so that the Earth is at coordinate (0, 1). In this normalisation the constant *k* corresponds to the fraction of the distance to Earth that the alien civilisation has reached so far.

[37] For instance, the shape of the eventual frontier between expanding civilisations that start at different times is given by the intersections of their future travel cones.

[38] As suggested by Olson (2016).

If we were to observe such a civilisation, it may have very different consequences to an observation with traditional SETI. This is because it may be millions or billions of years before we could interact. Or, if they are sufficiently distant and slow that they have not yet reached our affectable universe, we would *never* be able to interact. Despite this, the observation would still have transformative effects upon us, filling us with awe at the scale of their accomplishments, humility at how small we still are, and ambition to achieve our own ends across a cosmic scale.

If we did detect an expanding civilisation, it may be able to reach us more quickly than the distances would first suggest. For it is not just the apparent shape that is distorted, but also the speed at which this shape grows. *Figure 13* involved a civilisation that actually expands at 50% of the speed of light. But it would appear to expand towards us at the full speed of light.[39] We can see this from the spacetime diagram: the time interval between the moment the civilisation could first be detected and the moment its travel cone intercepts the Earth, is exactly the time interval during which the blue light ray travels the full distance in the opposite direction. If they were expanding any faster than 50% the speed of light, then the expansion wave would appear to come even faster than the speed of light (even more light years per year). As the expansion speed approaches 100% of the speed of light, the apparent expansion towards us would become instantaneous — in other words, the time interval between our detecting its origins billions of light years away and it arriving on our doorstep would approach zero.[40]

A civilisation that wished to remain unnoticed by others could take advantage of this effect, by delaying any activity that would be visible across cosmological distances. If it delays detectable activity in each galaxy until the time when light from that galaxy could no longer exit the final settleable region, then the civilisation would be invisible to other civilisations until it reaches their own galaxies (see *Figure 14*).

---

[39] There is a possibility for confusion here, as there are two senses of what it might mean to appear to approach at the speed of light. I mean this in the sense of coming closer to us by one light year per year (with appropriate adjustments for the expansion of space), so that it appears to approach at the actual speed of light. In contrast, something actually moving at the speed of light would appear to cross that distance instantaneously.

[40] This becomes less mysterious if we use their initial shape to compute their actual current extent and travel speed. What is really happening is not that a quickly travelling civilisation approaches the observer at beyond the speed of light, but that it is much closer than it appears, so has less far to travel before reaching the observer.

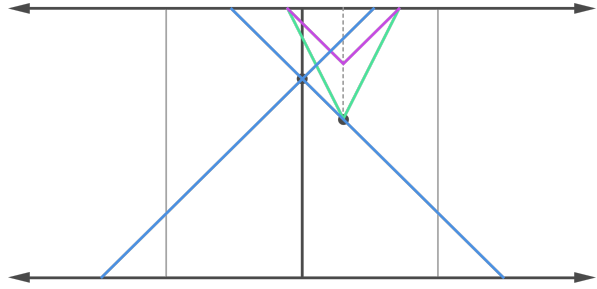

*Figure 14.* A modification of *Figure 13* in which the alien civilisation still expands according to the green travel cone, but in each galaxy it only commences activities that are detectable from afar at a later time, marked in purple. This would prevent it being observed from a distance — observers could only detect it from within.

The delay in some activities would have costs to the civilisation, but these may not be too high — especially if the civilisation is patient (not minding when it achieves its ends) and if the starlight wasted while they delay is only a small fraction of the harnessable energy of each galaxy. In this case, the benefits of going unnoticed could outweigh the costs, such that these vast civilisations would remain undetectable until the moment they collide.

**What if ΛCDM is wrong?**

All of the preceding discussion on the edges of the universe and their implications has assumed the widespread ΛCDM model of cosmology. What if this is wrong? One of the biggest uncertainties in cosmology is the nature of dark energy — the property of our universe that gives rise to the accelerating expansion that we have observed. ΛCDM has a very simple model of dark energy: it assumes that it is constant across space and time. Other models allow dark energy to vary. How would this change these results?

A useful way of categorising the possibilities concerns the value of an unknown parameter, $w$. This is the parameter in the 'equation of state' for a perfect fluid, and is equal to its pressure divided by its energy density. The regular matter in the universe has $w = 0$. Relativistic matter has $w = 1/3$. ΛCDM models dark energy as a cosmological constant, which corresponds to $w = -1$. This causes expansion which becomes exponential in the long run, leading to the results we have discussed. If dark energy is better modelled by a value of $w$ between $-1$ and $-1/3$, then expansion won't become exponential, but will still continue to accelerate, leading to roughly similar results — in particular that only a finite number of galaxies are ever affectable. However, if $w$ is actually less than $-1$, or greater than $-1/3$ (or if the entire appearance of an accelerating expansion is just some kind of measurement artefact), then the future of the universe will be remarkably different. Our current best estimates of $w$ are consistent with ΛCDM: putting it to within about 10% of $-1$, but the other models cannot yet be excluded.

If $w$ were below $-1$, then the scale factor would grow faster than an exponential. This would enable expansion to tear apart all 'bound' systems including clusters, galaxies, solar systems, planets, and even atoms. Furthermore, the scale factor would reach infinity in a finite time, meaning that by a particular year the proper distance between any pair of particles would become infinite. Presumably this moment would mark the end of time. This scenario is known as the 'Big Rip'. Thus in a universe with $w$ below –1, an ambitious civilisation may have much less time to achieve its ends, and its strategies may have to be much less patient than those I sketched earlier.

If $w$ were between –1/3 and 0, then the scale factor would merely grow sub-linearly, making it easier to travel between distant galaxies and removing the finite limit on the number of reachable galaxies. In the terms of the diagrams I have used, the shape of the path of a ray of light would no longer have a vertical asymptote and it would instead continue past arbitrarily many galaxies.

At one level, this scenario would appear to be extremely different for ambitious civilisations: they would be free to explore a truly unbounded part of the cosmos, allowing potentially unlimited resources and potentially infinitely long life. However, there is reason to suspect that this may really just mean larger but still finite amounts of resources and lifespan. One issue is that after a finite time it is widely suspected that most matter will have decayed, so that on arrival at almost all locations the only available resources would be occasional electrons, positrons, and low energy photons, separated by unimaginably vast distances. It may thus be impossible to harness these decayed resources using less energy than they would grant you.

Another issue is that in a large enough universe with no travel limits, a civilisation still wouldn't have a truly unlimited domain to itself. It would eventually encounter other independently arising civilisations with similar objectives. This is true no matter how low the probability of them arising on any particular planet (so long as this is not precisely zero). The growing spheres of influence of these civilisations would eventually meet one's own on all sides, dividing the infinite resources up into a finite parcel for each of the infinitely many civilisations. While there are mathematical ways of arranging an infinite set of resources so that infinitely many parties each get an infinite share (*c.f.* Hilbert's Hotel), it is not clear that any of these are practical in our universe. They typically involve picking out a particular location as the origin of a coordinate system and assigning each party's share in reference to that location. This may be impossible to arrange in our universe which appears to have no privileged centre or orientation, and only a finite speed of communication between the parties.

However neither of these problems would seem to prevent civilisations each gaining access to a region of empty space which grows without bound in terms of proper size and yet stays in causal contact. Thus if $w > -1/3$, it may just be possible to enact Dyson's scheme for eking out a finite amount of free energy into a literally infinite amount of computation and, perhaps, a truly infinite lifespan.[41]

---

[41] (Dyson 1979). Compare with my comments on the impossibility of this scheme within ΛCDM, in the section *Eras of the Universe*.

**SUMMARY**

We have used spacetime diagrams for a universe with accelerating expansion to distinguish a total of nine different spherical boundaries around us that could be said to be the edge of our universe. All except the Hubble volume capture a causal relationship between the centre and the other points, which can be neatly explained with spacetime diagrams. In doing so, we have seen that we cannot affect everything in the observable universe, but we *can* observe more than the (currently) observable universe if we wait longer or move from our current location.

In the second part, we briefly examined ways in which these spheres would affect the potential activities of an interstellar civilisation including setting relevant timescales and bounding their resources, information, and interactions.

Here is a summary of the main causal boundaries that we have discussed.

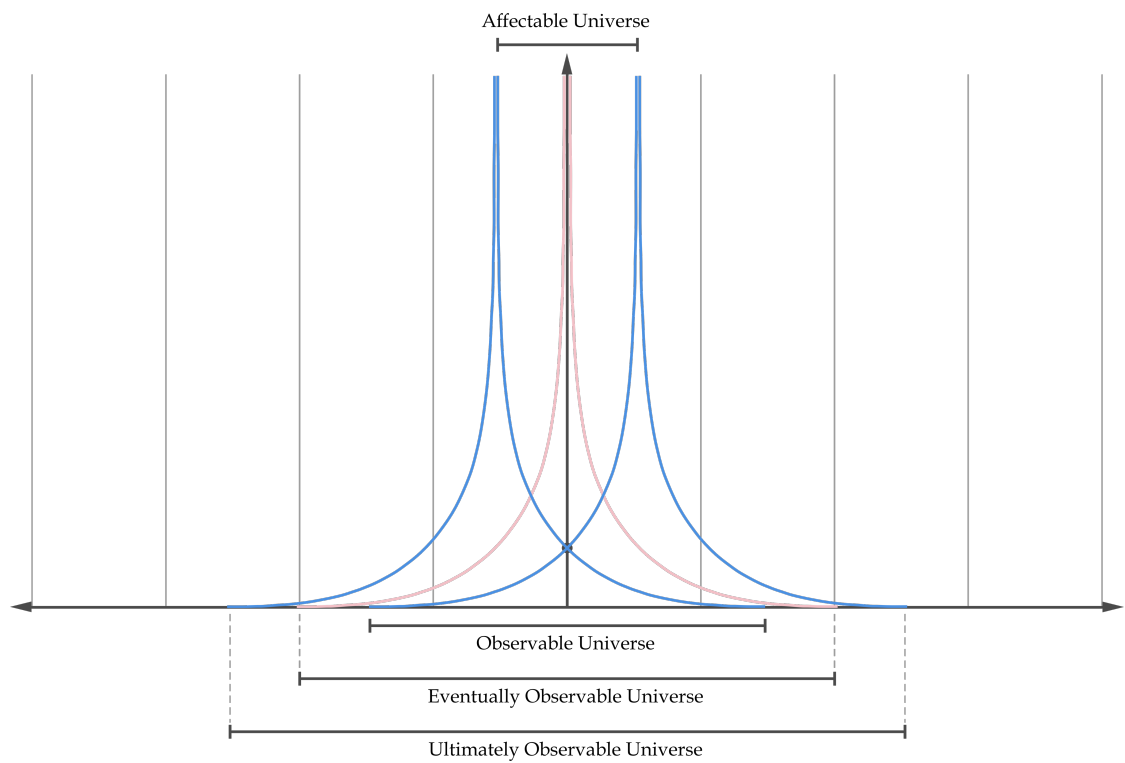

*Figure 7 (repeated).* The major causal regions of our universe.

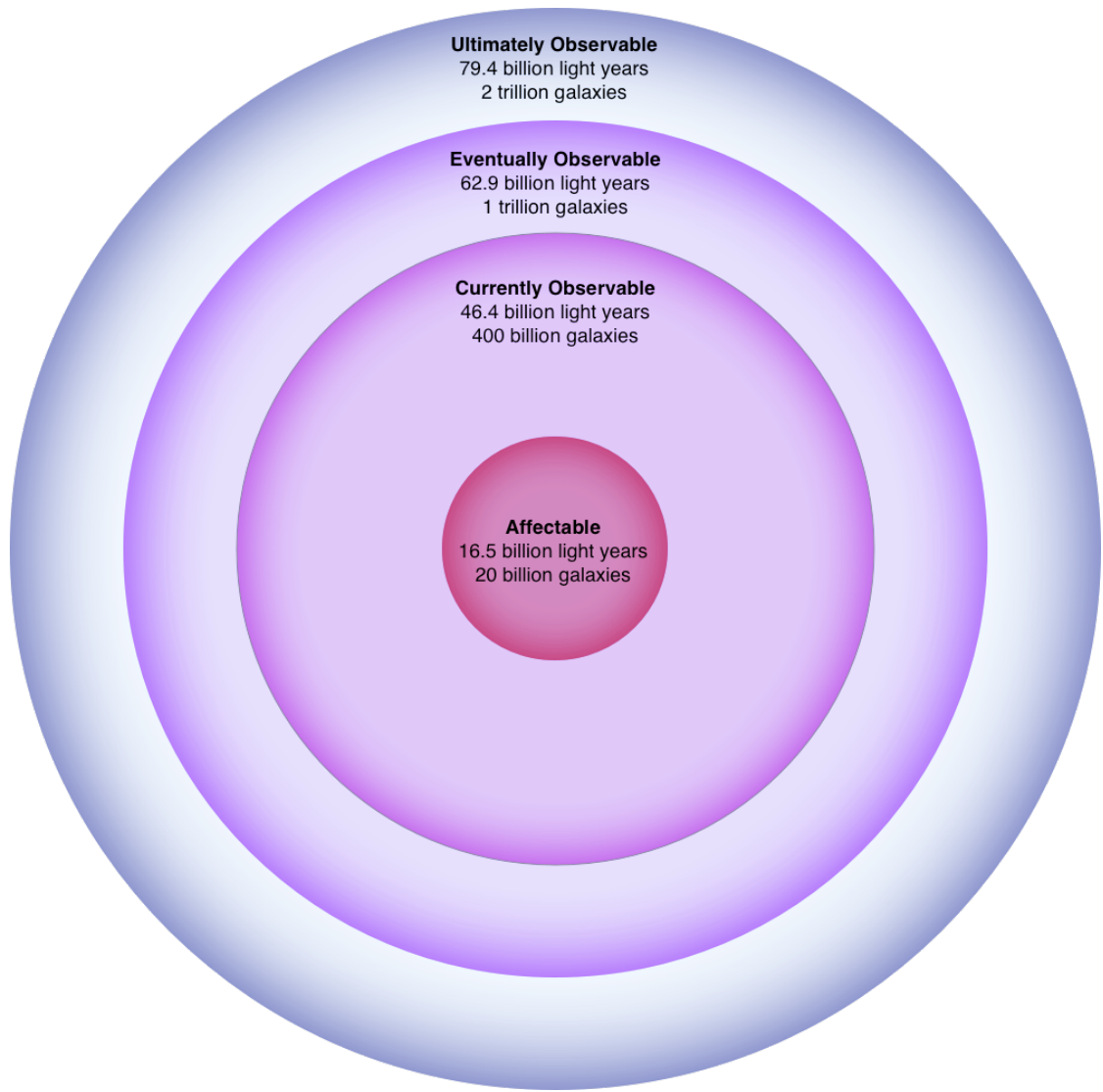

*Figure 15.* The most important causal boundaries, in proportion.

*Figure 16 (overleaf).* Zooming in from the scale of these causal boundaries to our own galaxy.

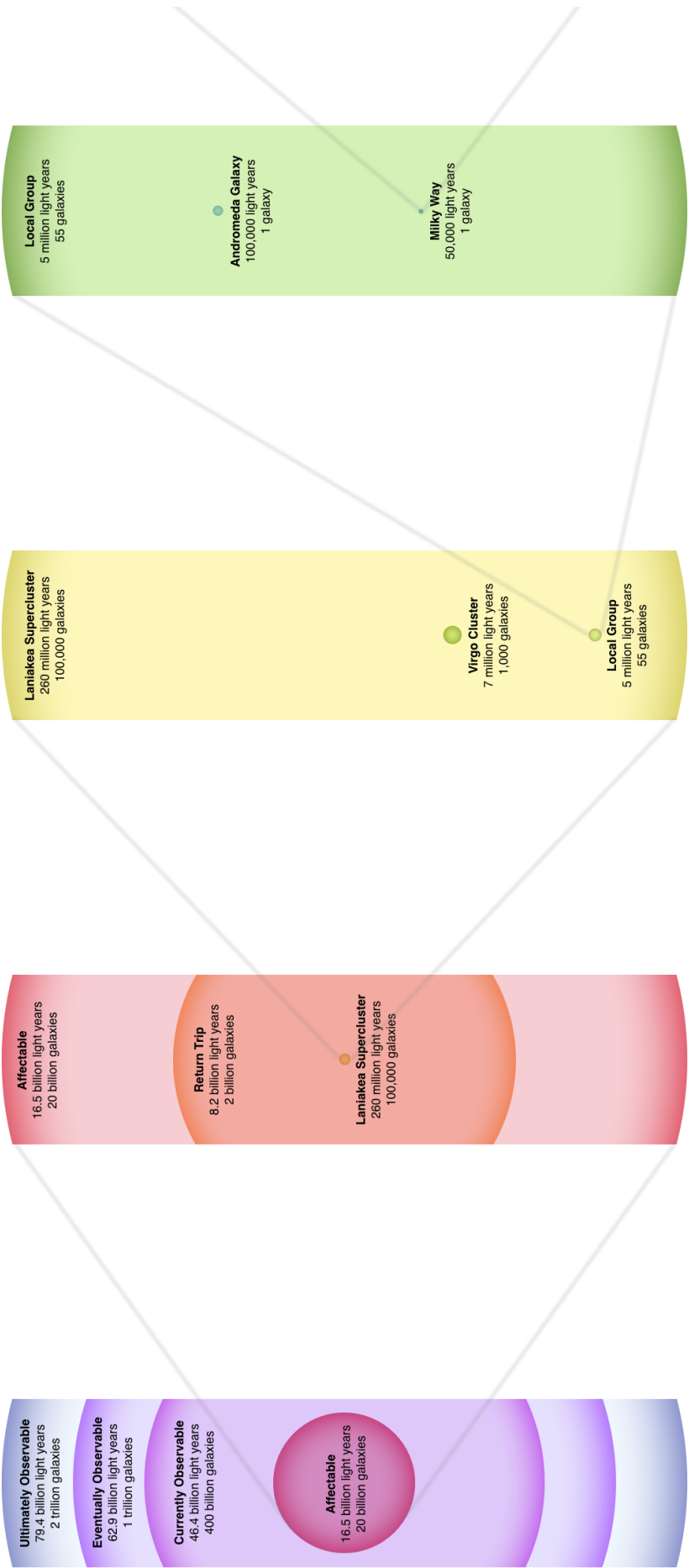

| | |
|---|---|
| $r_a(t) = D_\gamma - d_\gamma(t)$ | *Affectable Universe*   (16.5 billion light years) |
| | The part of the universe we can causally affect. |
| | The part of the universe whose current events will eventually be visible from Earth. |
| $r_o(t) = d_\gamma(t)$ | *Observable Universe*   (46.4 billion light years) |
| | The part of the universe that can causally affect us, now. |
| | The part of the universe we can currently see. |
| $r_{eo} = D_\gamma$ $= \lim_{t \to \infty} d_\gamma(t)$ | *Eventually Observable Universe*   (62.9 billion light years) |
| | The part of the universe that can eventually causally affect us here. |
| | The part of the universe we will eventually be able to see from here. |
| $r_{uo}(t) = D_\gamma + r_a(t)$ | *Ultimately Observable Universe*   (79.4 billion light years) |
| | The part of the universe that can eventually causally interact with us. |
| | The part of the universe we might eventually be able to see, with travel. |

And here are some spheres that are noticeably less fundamental than the above, but still worthy of note:

| | |
|---|---|
| $r(t) = r_a(t) / 2$ | —   (8.2 billion light years) |
| | The part of the universe we could theoretically get to and return. |
| | The largest completely casually connected region, in the sense that every point can observe and reach every other. |
| $r_H(t) = c / H(t)$ | *Hubble volume*   (14.4 billion light years) |
| | The part of the universe receding from us at less than light speed. |
| | (This appears to have no interesting properties relating to causality.) |
| $r(t) = r_a(t) + r_a(t)$ | —   (33.0 billion light years) |
| | The part of the universe whose current events we might eventually be able to see, with travel. |
| $r(t) = D_\gamma + r_o(t)$ | — (109.3 billion light years)<br><br>The part of the universe affected by events that have also affected us.<br><br>The part of the universe that can eventually be affected by events we can currently see. |
| $r = D_\gamma + D_\gamma$ | — (125.8 billion light years)<br><br>The part of the universe which can ever have any kind of causal connectedness to our location. |

We shall finish with an photograph from the Hubble Space Telescope's Extreme Deep Field. This image includes the most distant galaxies and protogalaxies we ever have seen. The largest galaxies in the image are as close as 6 billion light years, while the smallest dots are protogalaxies as far away as 30 billion light years. This image thus spans the edge of our affectable universe, with most of the places it shows being forever beyond our reach.

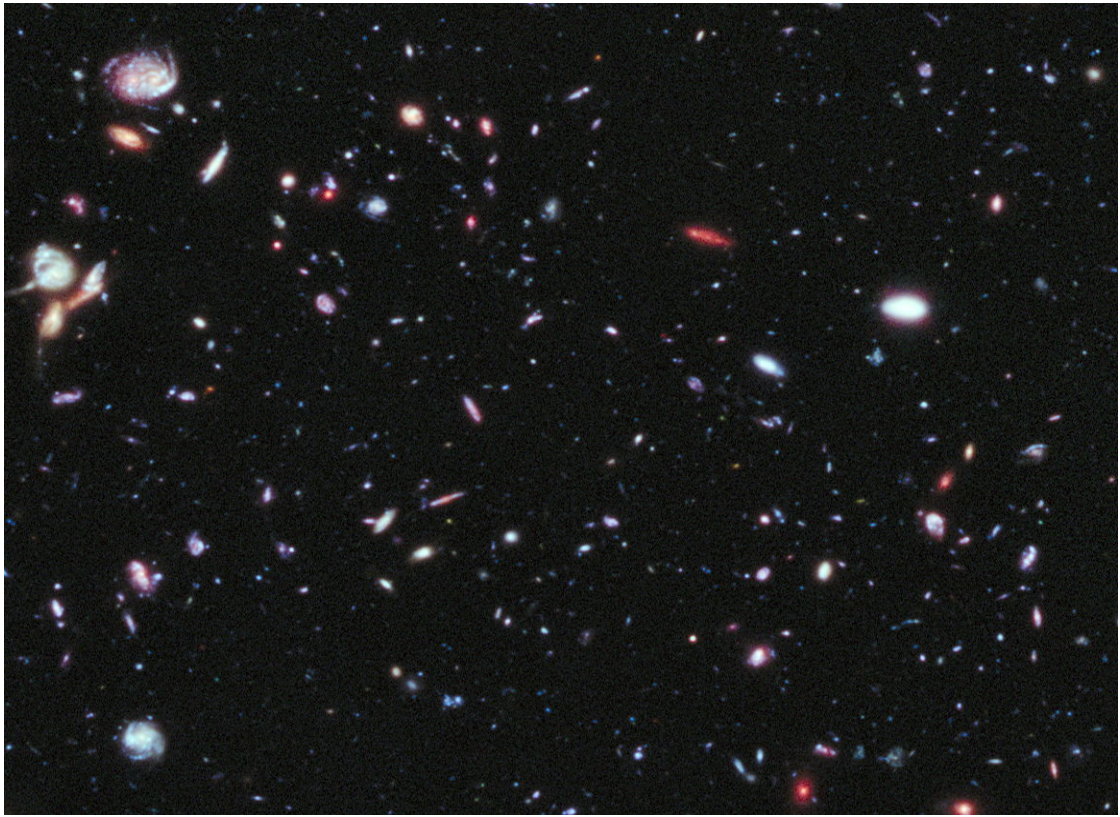

**APPENDIX**

The mathematics of these edges of the universe is almost entirely determined by the evolution of the scale factor over time, $a(t)$. However, there is no closed form expression for $a(t)$, which makes it difficult to work algebraically with or to calculate. Fortunately, it is a one-to-one function and its inverse can be expressed as:

$$t(a) = \frac{1}{H_0} \int_0^a \frac{a' \mathrm{d}a'}{\sqrt{\Omega_{R,0} + \Omega_{M,0} a' + \Omega_{K,0} a'^2 + \Omega_{\Lambda,0} a'^4}}$$

where:

| | | |
|---|---|---|
| $H_0$ | $\approx 1/(14.4 \text{ billion years})$ | Current value of the Hubble parameter |
| $\Omega_{R,0}$ | $\approx 0.000098$ | Current density of radiation |
| $\Omega_{M,0}$ | $\approx 0.308$ | Current density of matter |
| $\Omega_{K,0}$ | $\approx 0$ | Current curvature of space |
| $\Omega_{\Lambda,0}$ | $\approx 0.692$ | Current density of dark energy |

One can numerically evaluate this formula across a vast range of values of $a$ using a spreadsheet, then use the resulting correspondences between $t$ and $a$ values as a lookup table for $a(t)$. If trying this, I recommend starting $a$ at 0.000001 then increasing it in compounding steps of about half a percent, until it reaches 1,000,000 (6,000 rows down). Then in each row, calculate the corresponding values of $t$, $d_\gamma(t)$, $r_a$, and $z$. See *Table 3* for some key values in such a table, computed using the parameters and method above.

Let $d_\gamma(t)$ be the distance light has been able to travel by time $t$. It is related to the scale factor by the following equation:

$$d_\gamma(t) = \int_0^t \frac{c}{a(t')} dt'$$

We can then define $D_\gamma$ as the ultimate distance light can travel (roughly 62.9 billion light years).

Distances to other galaxies are often given via the amount of redshift that their light undergoes on its journey to us. The amount of redshift, $z$, is easily convertible to the scale factor at the time the light was emitted using the equation:

$$a(t) = \frac{1}{1+z}$$

| t | a(t) | $d_\gamma(t) = r_o$ | $r_a$ | Note |
|---|---|---|---|---|
| 1 *y* | 0.0000015 | 790 *kly* | 62.93 *Gly* | |
| 380 *ky* | 0.00091 | 0.91 *Gly* | 62.0 *Gly* | *Universe becomes transparent* |
| 10 *My* | 0.0072 | 3.6 *Gly* | 59.3 *Gly* | |
| 100 *My* | 0.033 | 8.5 *Gly* | 54.4 *Gly* | |
| 1 *Gy* | 0.15 | 19.3 *Gly* | 43.6 *Gly* | |
| 13.8 *Gy* | 1 | 46.4 *Gly* | 16.5 *Gly* | *Present day* |
| 150 *Gy* | 2754 | 62.92 *Gly* | 0.0063 *Gly* | *Era of isolation begins* |
| ∞ | ∞ | 62.93 *Gly* | 0 *Gly* | *Limit of infinite time* |

*Table 3.* A sample of useful numerical values for cosmological factors and distances over time.